\documentclass[a4paper,10pt,twocolumn,preprint,3p]{elsarticle}

\usepackage{booktabs} % for much better looking tables
\usepackage{color}
\usepackage{bm}
\usepackage[utf8]{inputenc} % set input encoding (not needed with XeLaTeX)

\newcommand{\bfD}{{\bf D}}%
\newcommand{\bfI}{{\bf I}}%
\newcommand{\bfu}{{\bf u}}%
\newcommand{\bftau}{\boldsymbol{\tau}}%
\newfont{\tenbfit}{cmbx10}%

\newfont{\tenbbb}{msbm10}%
\newfont{\svnbbb}{msbm8}%
\newcommand{\half}{{\textstyle{\frac{1}{2}}}}

\newcommand{\Grad}{\hbox{\rm grad}\mskip2mu}
\newcommand{\Div}{\hbox{\rm div}\mskip2mu}

\newcommand{\trans}{\mskip-2mu\scriptscriptstyle\top\mskip-2mu}

\newcommand{\lj}{\mbox{$[\kern-0.1478125em[$}}
\newcommand{\rj}{\mbox{$]\kern-0.1478125em]$}}
\newcommand{\la}{\mbox{$\langle\kern-0.2325em\langle$}}
\newcommand{\ra}{\mbox{$\rangle\kern-0.2325em\rangle$}}
\newcommand{\Blj}{\mbox{$\Big[\kern-0.275em\Big[$}}
\newcommand{\Brj}{\mbox{$\Big]\kern-0.275em\Big]$}}
\newcommand{\Bla}{\mbox{$\Big\langle\kern-0.425em\Big\langle$}}
\newcommand{\Bra}{\mbox{$\Big\rangle\kern-0.425em\Big\rangle$}}

\newcommand{\zed}{{\bf 0}}

\journal{}
\begin{document}

\begin{frontmatter}

\title{Performance of eddy-viscosity turbulence models for predicting swirling pipe-flow: Simulations and laser-Doppler velocimetry}

\author[dod]{Diego del Olmo D\'iaz} 
\author[dfh]{Denis F.\ Hinz\corref{cor}}
\ead{dfh@kamstrup.com}

\cortext[cor]{Corresponding author}

\address[dod]{Université de Picardie Jules Verne, Chemin du Thil, 80000 Amiens, France}
\address[dfh]{Kamstrup A/S, Industrivej 28, Stilling, 8660 Skanderborg, Denmark}

\begin{abstract}
We use laser-Doppler velocimetry (LDV) experiments and Reynolds-averaged Navier--Stokes (RANS) simulations to study the characteristic flow patterns downstream of a standardized clockwise swirl disturbance generator.
After quantifying the impact of the mesh size, we evaluate the potential of various eddy-viscosity turbulence models in providing reasonable approximations with respect to the experimental reference.
The choice of turbulent models reflects current industry practice.
Our results suggest that models from the {$k$-$\epsilon$} family are more accurate in predicting swirling flows than models from the {$k$-$\omega$} family.
For sufficiently resolved meshes, the realizable {$k$-$\epsilon$} model provides the most accurate approximation of the velocity magnitudes, although it fails to capture small-scale flow structures which are accurately predicted by the standard {$k$-$\epsilon$} model and the RNG~{$k$-$\epsilon$} model. 
Throughout the article, we highlight practical guidance for the choice of RANS turbulence models for swirling flow.
\end{abstract}

\end{frontmatter}

%%%%%%%%%%%%%%%
\section{Introduction}
\label{sec:intro}
%%%%%%%%%%%%%%%

Swirling flows are ubiquitous in many industrial applications including furnaces, cyclone separators, heat exchangers, and turbines. Yet, efficient numerical prediction with Reynolds-averaged Navier--Stokes (RANS) approximations is challenging and large-eddy simulations (LES) are still prohibitively expensive for practical applications. For strongly rotating flows, various standard assumptions used in RANS turbulence closures are not expected to hold, as discussed, for example, by Jakirli\'c et al.~\cite{Jakirlic2002}. Consequently, the performance of conventional RANS modeling approaches for predicting swirling flow remains elusive and requires further verification and validation with experiments.

Kitoh~\cite{Kitoh1991} performed experiments with turbulent swirling flow in a straight pipe where the swirl component is generated through guide vanes with variable vane angle and a bell-shaped cone at the center of the swirl generator. Based on the characteristic tangential velocity distribution, Kitoh~\cite{Kitoh1991} categorized the flow field into three regions: wall, annular, and core. Due to the streamline curvature, skewed shear directions in the annular region, and the resulting anisotropy, Kitoh~\cite{Kitoh1991} concludes that Reynolds stress models seem more promising than eddy-viscosity models in providing accurate predictions of swirling flows. This conclusion confirms earlier simulation results of Kobayashi and Yoda~\cite{Kobayashi1987} and is also supported by subsequent numerical studies. For example, simulations of Hirai et al.~\cite{Hirai1988} show a better performance of Reynolds stress models over the standard {$k$-$\epsilon$} model and a modified {$k$-$\epsilon$} model for the prediction of the laminarization phenomenon in swirling pipe-flow. Similarly, \'Co\'ci\'c et al.~\cite{Cocic2014} report that two-equation turbulence models including RNG~{$k$-$\epsilon$}, Lauder Sharma {$k$-$\epsilon$}, and {$k$-$\omega$}~SST fail to predict experimental data, whereas the Launder--Gibson Reynolds stress model and the Speziale--Sarkar--Gatski model are found to provide better agreement with experiments.

In contrast, Chen et al.~\cite{Chen1999} find that the differential Reynolds stress model proposed by Launder et al.~\cite{Launder1975} provides quantitatively inaccurate results for predicting tangentially injected swirling pipe flows, reaching only qualitative agreement with experimental results. Further, several workers successfully applied turbulence models of the {$k$-$\epsilon$} family to model swirling flow. For example, Parchen and Steenbergen~\cite{Parchen1998} studied turbulent swirling pipe-flows for gas-metering applications and reported an acceptable behavior of the {$k$-$\epsilon$} model, which provided a better approximation of the experimental values compared to an algebraic slip model. Escue and Cui~\cite{Escue2010} found a superior performance of the RNG~{$k$-$\epsilon$} model over a Reynolds stress model for moderate swirl intensities in 2D simulations when compared to experimental results. Petit et al.~\cite{Petit2011} successfully applied a {$k$-$\epsilon$} model for unsteady simulations of a swirl generator, achieving  good agreement in averaged velocity profiles. In view of the available results, a general consensus regarding the performance of eddy-viscosity closures for swirling flows appears to be lacking. 

%%%%%%%%%%%%%%%
\subsection{Swirling flows in flow metering}
\label{sec:Sflow}
%%%%%%%%%%%%%%%

\begin{figure}[t!]
\centering

\includegraphics[height=0.15
\textwidth, trim=0.2cm 0.2cm 0.2cm 0.2cm, clip=true] {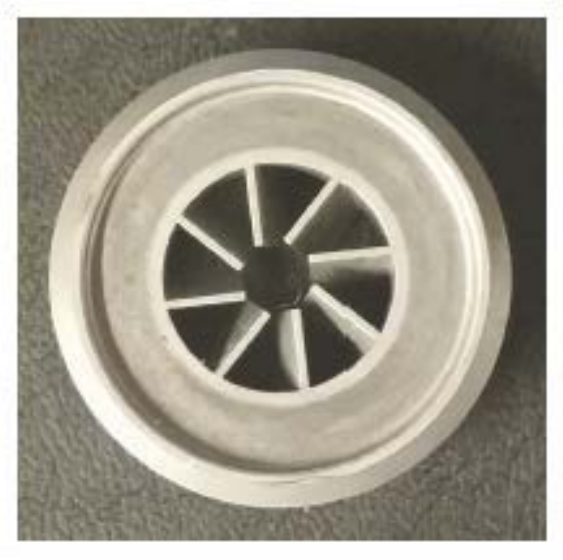}\put(-85,70){(a)} 
$\qquad$
\includegraphics[height=0.15
\textwidth, trim=0cm 0cm 0cm 0cm, clip=true] {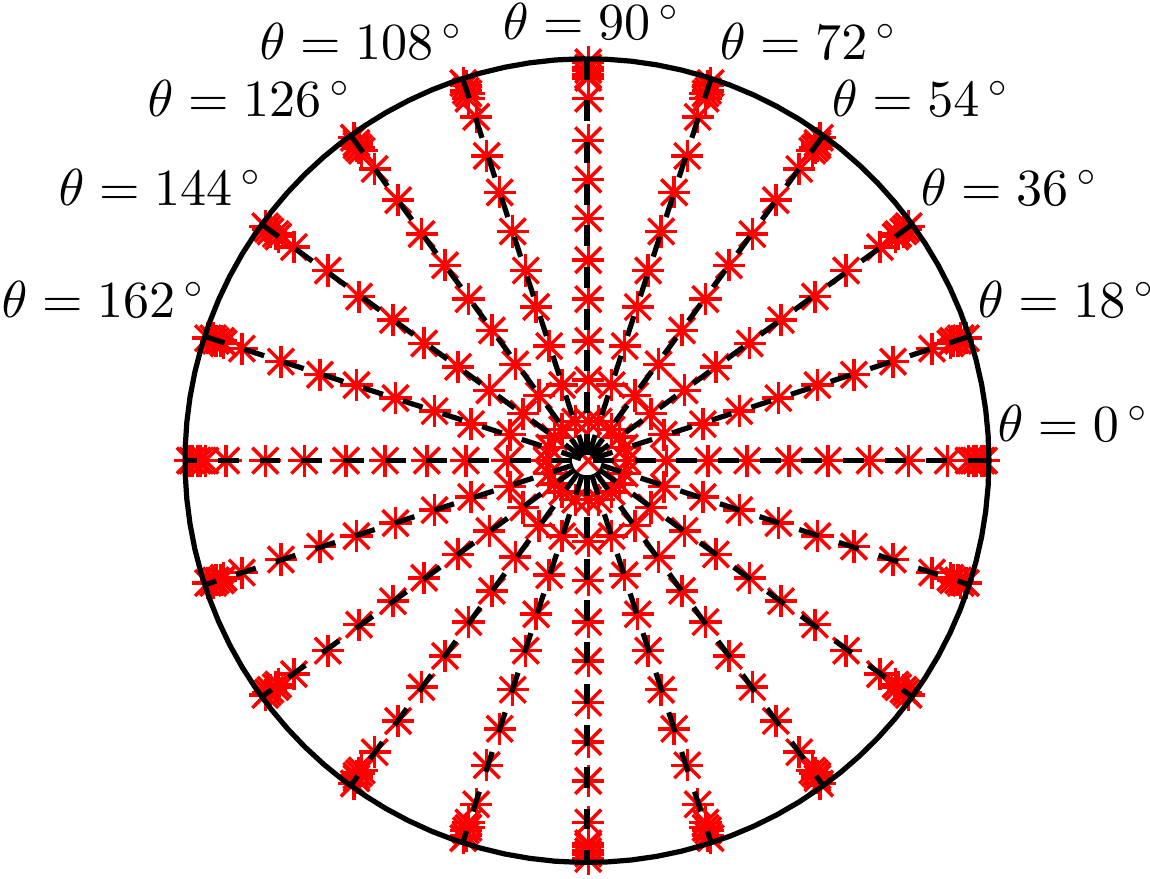}\put(-100,70){(b)} 
\vspace{0.2cm}
\includegraphics[height=0.15
\textwidth, trim=0cm 0cm 0cm 0cm, clip=true] {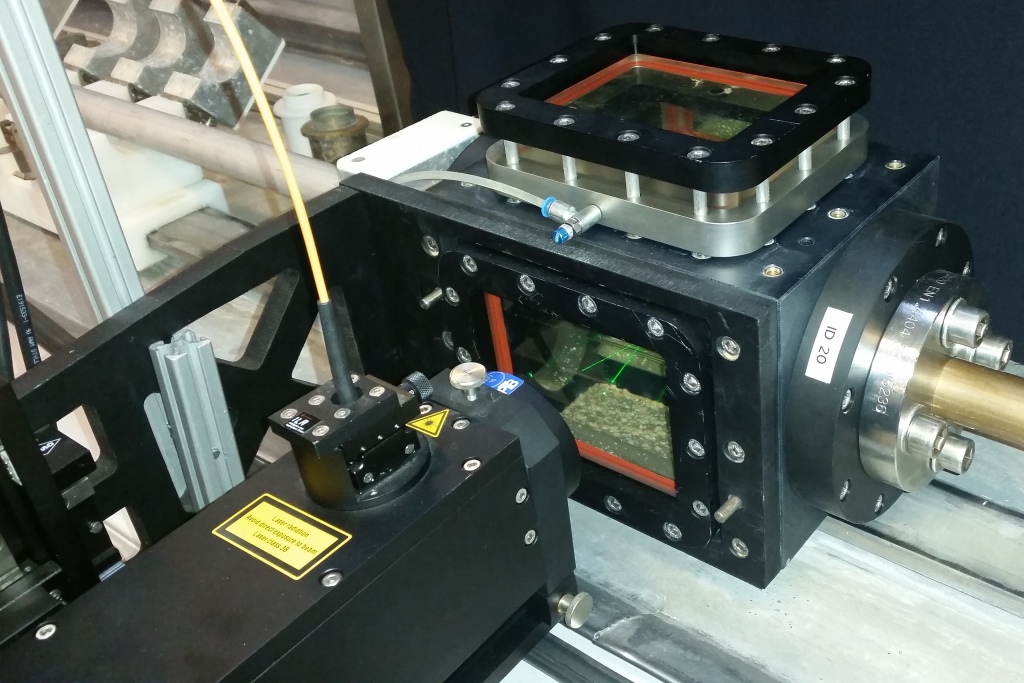}\put(-120,70){(c)} 

\caption{(a) Standarized swirl disturbance generator according to {EN~ISO~4064-2:2014}~\cite{ENSO4064} and {OIML~R~49-2:2013}~\cite{OIML2013}. (b) Measurement grid for the LDV experiments. (c) LDV probe and optical access from ILA/Optolution.}
\label{fig:swirlGrid}
\end{figure}

In this article, we study the swirling flow generated by a standardized clockwise swirl disturbance generator that is used for testing commercial flow meters (Figure~\ref{fig:swirlGrid} (a)). Commercial water, heat and cooling meters may be exposed to disturbed inflows when they are in operation, since the space available in realistic installations is often limited and prohibits the installation of a straight pipe long enough to create fully developed flow conditions upstream of the meter. Common flow disturbances induced by several standard installations such as bent pipes or valves may compromise the accuracy of meter readings. Consequently, it is necessary to assess the robustness of new products with respect to disturbed flow conditions. The standards {EN~ISO~4064-2:2014}~\cite{ENSO4064} and {OIML~R~49-2:2013}~\cite{OIML2013} for the type-approval of commercial water meters and {EN~1434-4:2007}~\cite{EN1434} for the type-approval of commercial heat and cooling meters include standardized tests with artificially generated disturbed flows that are meant to emulate disturbed flow conditions in realistic installations. One of these artificially disturbed flows is generated by a standardized clockwise swirl disturbance generator (Figure~\ref{fig:swirlGrid} (a)), which emulates the flow structures induced by a double-bent pipe installation. (For a study of flow structures induced by double-bent pipes see, for example, Mattingly and Yeh~\cite{Mattingly1991}.)

A detailed analysis of flow patterns generated by such swirl disturbance generators is not part of the standard type-approval procedure, but it is a key factor in enabling a meaningful interpretation of type-approval tests. Further, the detailed analysis of swirling flows has potential to assist the development process of robust meter designs and the associated testing and verification facilities. The swirling flow field generated with equipment according to {EN~ISO~4064-2:2014}~\cite{ENSO4064} and {OIML~R~49-2:2013}~\cite{OIML2013} is expected to exhibit similar features as the flow fields in the studies discussed in Section~\ref{sec:intro}. For example, Eichler and Lederer~\cite{Eichler2015} conducted stereoscopic particle image velocimetry (SPIV) measurements downstream of a standardized DN80 swirl disturbance generator, showing that the swirling flow is maintained up to $87.0D$ downstream. Similarly, Tawackolian~\cite{Tawackolian2013a} provides experimental and numerical results using the {$k$-$\omega$} model for a DN80 swirl generator and a comparison of the flow patterns produced by the swirl generator against those after bent pipes. While these studies provide valuable insight for large pipe diameters, no results appear to be available for smaller pipe diameters. Since the standardized parts for different diameters according to {EN~ISO~4064-2:2014}~\cite{ENSO4064} and {OIML~R~49-2:2013}~\cite{OIML2013} are not exactly self-similar, the scalability of available results to smaller diameters remains unclear.

The present work aims to asses the potential of RANS simulations in providing accurate predictions of the flow patterns downstream of a standardized swirl disturbance generator. In view of the discussion in Section~\ref{sec:intro}, available practical guidance for choosing turbulence models is ambiguous. To elucidate modeling choices for swirling pipe-flow, we study different numerical setups and compare them against experimental results obtained from laser-Doppler velocimetry (LDV) experiments. Our choice of turbulence models includes some of the most popular mainstream models currently used in industry. The aim is to identify best practices to achieve realistic and useful simulation results that can be used in practical applications such as flow meter testing and development. Additionally, we discuss our results in the context of available recommendations regarding the modeling of swirling flows. Within the study of different turbulence models, we also provide a detailed assessment of the impact of the mesh size and the length of the computational domain.

%%%%%%%%%%%%%%%
%%%%%%%%%%%%%%%

%%%%%%%%%%%%%%%
\section{Materials and methods}
\label{sec:mat_methods}
%%%%%%%%%%%%%%%

We use a standardized clockwise swirl disturbance generator according to {EN~ISO~4064-2:2014}~\cite{ENSO4064} and {OIML~R~49-2:2013}~\cite{OIML2013} with an inner diameter of $D=15.0 \, \rm{mm}$.
The working fluid is water and both simulations and experiments are realized with Reynolds number $\rm{Re}=4.0\cdot 10^4$, where $\rm{Re}$ is based on the volumetric velocity $w_{\rm vol}$ and the pipe diameter $D$, such that

\begin{equation}\label{eq:Re}
{\rm Re}= \frac{ w_{\rm vol} D}{ \nu},
\end{equation}
and
\begin{equation}\label{eq:wVol}
w_{\rm vol}= \frac{Q}{A},
\end{equation}
with $\nu$ the kinematic viscosity of the water, $Q$ the volumetric flow rate, and ${A=\frac{\pi}{4}D^2}$ the area of the pipe cross-section.

%%%%%%%%%%%%%%%
\subsection{Experiments}
\label{sec:exp}
%%%%%%%%%%%%%%%

All experiments are performed on a calibration flow bench with an expanded uncertainty  of $0.30\,\%$ (coverage factor $k=2.0$) for measurements against weight with test-volumes of $3\, \rm{liter}$ to $100 \, \rm{liter}$. The flow bench is equipped with three magneto-inductive (MID) master meters, each one responsible for controlling the flow in a certain flow-rate interval. The experiments are realized with water at temperature $T=20.0 \, \rm{^\circ C}$, corresponding to a kinematic viscosity of ${\nu = 1004.79\cdot 10^{-9}\, \rm{m^2/s}}$. During measurements, the volume flow-rate $Q$, the water temperature $T$, and the pressure $p$ are stabilized through PID controlled feedback loops. We verify the stability of $Q$, $T$, and $p$ by logging data from the master meters and the corresponding temperature and pressure sensors. Let $\overline Q_M$ denote the time-averaged master flow-rate and let $\sigma_{Q_M}$ denote the associated standard deviation providing a measure for the stability of the flow-rate. We find $\sigma_{Q_M}/\overline Q_M \approx 0.25\, \%$ and that the master meter signal has the characteristics of random white noise, which confirms that there is no preferred timescale and no low frequency disturbances that might bias the long-time accuracy of LDV measurements.  

We perform measurements with a commercial LDV system (ILA/Optolution) in and industrial flow laboratory. The flow is seeded with neutrally buoyant silver coated hollow glass beads to improve the LDV signal. To ensure a fully developed flow profile upstream, the swirl generator is located after more than $100.0D$ of straight pipe and the measurement section is located $12.0D$ downstream from the swirl disturbance generator. For the present study, we focus on discussing the flow patterns in the near-field until $12.0D$ downstream from the swirl disturbance generator, since this is the flow that will enter a flow meter in a type-approval test. The measurement grid comprises 241 measurement points as shown in Figure~\ref{fig:swirlGrid}~(b).

The amount of data acquired at each measurement point is determined by the choice of two experimental constraints: (I) the maximal number of single-point samples $n_{\rm{max}}$ and (II) a timeout $t_{\rm{max}}$  for each single-point measurement on the measurement grid. For the present measurements we choose 
\begin{equation}\label{eq:expCond}
\begin{array}{ll}
n_{\rm{max}}=10^3 \quad {\rm{and}} & \quad t_{\rm{max}}=60.0\, \rm{s}. 
\end{array}
\end{equation}
Axial velocity profiles are obtained through a collection of single-point measurements of the local axial mean velocity $\overline w$ over the measurement grid (Figure~\ref{fig:swirlGrid}). To determine $\overline w$, we use the estimator
\begin{equation}\label{eq:estimator}
\overline w = \frac{1}{n} \sum\limits_{i=1}^n w_i,
\end{equation}
with $w_i$ single-point samples of velocities and $n$ the number of samples. The associated dimensionless turbulence intensity is
\begin{equation}\label{eq:Tu01}
{\rm{Tu}} = \frac{\sigma_w}{\overline w},
\end{equation}
where 
\begin{equation}\label{eq:sDevDef01}
\sigma_w = \left( \frac{1}{n-1} \sum\limits_{i=1}^n (w_i - \overline w)^2  \right)^
{1/2}
\end{equation}
is the standard deviation of samples $w_i$. To estimate the reliability of the estimator~\ref{eq:estimator} at each spatial measurement point, we determine the associated standard error
\begin{equation}\label{eq:sdErr}
\sigma_{\overline w} = \sigma_w / \sqrt{n}.
\end{equation}
The standard error~\ref{eq:sdErr} provides an uncertainty estimate for the estimator of the mean velocity~\ref{eq:estimator}. The mean velocity components $\overline u$ in $x$ direction and $\overline v$ in $y$ direction are determined analogously. To obtain all velocity components, three consecutive LDV measurements are required. Notice that the velocity vector $\overline \bfu = (\overline u, \overline v, \overline w)^{\trans}$ is the experimental equivalent of the Reynolds-averaged velocity defined in Section~\ref{sec:sim}.

%%%%%%%%%%%%%%%
\subsection{Simulations}
\label{sec:sim}
%%%%%%%%%%%%%%%

The numerical simulations are performed using a Reynolds-averaged Navier--Stokes (RANS) approach. Let a superposed bar denote Reynolds-averaging, so that (see, for example Pope~\cite{Pope2000})
\begin{equation}\label{eq:ReynoldsAvg01}
\bfu = \overline \bfu + \bfu^\prime,
\end{equation}
where $\overline \bfu$ is the RANS velocity and $\bfu^\prime$ is the fluctuating velocity satisfying $\overline{\bfu^\prime}=\zed$. Applying Reynolds-averaging to the incompressible Navier--Stokes (NS) equations yields the RANS equations
\begin{equation} \label{eq:RANS}
\left.
\begin{array}{l}
\dot{\overline \bfu}= -\frac{1}{\rho}\Grad \overline p +\nu \triangle  \overline \bfu - \Div \bftau,
\\
\Div \bfu = 0,
\end{array}
\!\!
\right\}
\end{equation}
where a superposed dot denotes material time-differentiation following $\overline \bfu$, $\overline p$ denotes the averaged pressure, $\rho$ denotes the fluid density, and 
\begin{equation}\label{eq:Reynolds_stress}
\bftau = \overline{\bfu^\prime \otimes \bfu^\prime }
\end{equation}
is the Reynolds stress tensor. The closure problem in RANS modeling amounts to finding expressions for~\ref{eq:Reynolds_stress} without using the fluctuating velocity $\bfu^\prime$. The gradient-diffusion and the eddy-viscosity hypotheses assume that the deviatoric part of \ref{eq:Reynolds_stress} is proportional to the averaged stretching tensor $\bfD = \half ( \Grad \overline \bfu + (\Grad \overline \bfu )^{\trans})$ and yield the closure approximation
\begin{equation}\label{eq:gradDiffEddyVisc}
- \bftau + \frac{2}{3}k \bfI = \nu_{\rm T} \bfD,
\end{equation}
where $\nu_{\rm T}$ is the eddy-viscosity, $\bfI$ is the identity tensor, and 
\begin{equation}\label{eq:kinE}
k=\half \overline{|\bfu^\prime|^2} 
\end{equation}
is the turbulent kinetic energy. In view of~\ref{eq:RANS}, \ref{eq:gradDiffEddyVisc}, and~\ref{eq:kinE}, the RANS equations with the gradient-diffusion hypothesis and the eddy-viscosity hypothesis are
\begin{equation} \label{eq:RANS_02}
\left.
\begin{array}{l}
\dot{\overline \bfu}=(\nu+\nu_{\rm T}) \triangle \overline \bfu -\frac{1}{\rho}\Grad (\overline p+\frac{2}{3}\rho k),
\\
\Div \bfu = 0.
\end{array}
\!\!
\right\}
\end{equation}
Notice that~$k$ can be absorbed into a modified averaged pressure. 

In this article, we consider various two-equation eddy-viscosity models (Table~\ref{t:modelRef}) that determine $\nu_{\rm T}$ through solving two additional model transport equations. The first additional  model transport equation is for the turbulent kinetic energy $k$ and the second additional model transport equation depends on the model family. For models from the {$k$-$\epsilon$} family, the second additional transport equation is for the turbulent dissipation 
\begin{equation}\label{eq:turbDiss}
\epsilon = 2 \nu \overline{\bfD^\prime : \bfD^\prime}
\end{equation}
where $\bfD = \half ( \Grad \overline{\bfu^\prime} + (\Grad \overline{\bfu^\prime} )^{\trans})$ is the fluctuation stretching tensor. Similarly, for models from the {$k$-$\omega$} family, the second additional transport equation is for the turbulent frequency $\omega = \epsilon /k$. Despite the vast selection of available turbulence models, we focus on popular mainstream models that are commonly used in industrial application. More exotic models specifically developed for swirling applications might well yield improvements in performance with the downside of giving up universality, which can be problematic in industrial applications. For a discussion of model performance, also see Section 5.

%%%%%%%%%%%%%%%
\subsubsection{Numerical solution}
%%%%%%%%%%%%%%%

We use the open-source CFD code OpenFOAM~\cite{OpenFOAM} and assess five different two-equation eddy-viscosity models, as summarized in Table~\ref{t:modelRef}. Both convective and diffusive terms are discretized with a second order Gaussian linear scheme. The near-wall region is modeled with enhanced wall functions available in  the standard OpenFOAM distribution. A SIMPLE algorithm is used for the coupling between the pressure and the velocity equations. The pressure equation is solved with a geometric agglomerated algebraic multigrid (GAMG) method, and all other equations are solved with the \emph{smoothSolver} solver of OpenFOAM using a Gauss--Seidel smoother.

The computational domain has a total length of $400.0 \, \rm{mm}$ with the swirl disturbance generator placed at the center. Consequently, the distance from both the inlet and outlet to the swirl generator is approximately $13.0D$. We use hex-dominant unstructured meshes generated with the OpenFOAM meshing tool \emph{snappyHexMesh}. All meshes are generated with the mesh quality constraints summarized in Table~\ref{t:meshQuality} in~\ref{sec:meshQ}.
%
%%%%%%%%%%%%%%%
\begin{table}[t!]%[H] add [H] placement to break table across pages
\centering
\footnotesize
%\small
\caption{Turbulence models and references of the specific model formulations used in the implementation in OpenFOAM.}
\label{t:modelRef}
\begin{tabular}{llllp{2cm}p{2cm}lp{2cm}}
\toprule
        Model    &Main reference(s)	\\ 
\hline
standard {$k$-$\epsilon$}  		&Launder and Spalding~\cite{Launder1974}\\
RNG~{$k$-$\epsilon$}  		&Yakhot et al.~\cite{YakhotRNG} \\
realizable {$k$-$\epsilon$}  		&Shih et al.~\cite{Shith1995} \\
{$k$-$\omega$}  			&Wilcox~\cite{Wilcox1988} \\
{$k$-$\omega$}~SST  		&Menter and Esch~\cite{Menter2001} \& Hellsten~\cite{Hellsten1998} \\ %%and Hellsten 
\bottomrule
\end{tabular}
\end{table}
%%%%%%%%%%%%%%%

%%%%%%%%%%%%%%%
\section{Results}
\label{sec:results}
%%%%%%%%%%%%%%%

%%%%%%%%%%%%%%%
\begin{figure*}[t!]
\centering
%\graphicspath{ {./pictures//} }
%
$\qquad$
\includegraphics[width=0.40\textwidth, trim=0cm 0cm 0cm 0cm, clip=true] {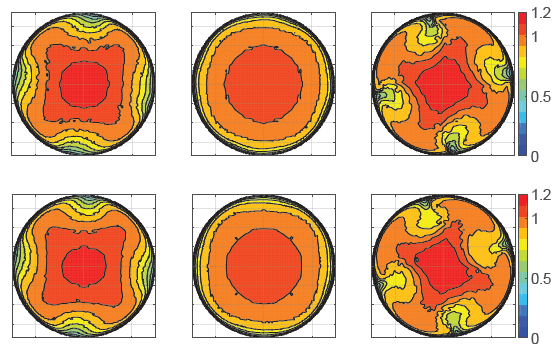}
\put(-190,130){(a)} 
\put(-178,118){RNG~{$k$-$\epsilon$}} 
%\put(-95,118){rea {$k$-$\epsilon$}} 
\put(-125,118){realizable {$k$-$\epsilon$}} 
\put(-55,118){{$k$-$\omega$}~SST} 
\put(-195,75){{\rotatebox{90}{short}}} 
\put(-195,15){{\rotatebox{90}{long}}} 
\hspace{0.25cm}
\includegraphics[width=0.40\textwidth, trim=0cm 0cm 0cm 0cm, clip=true] {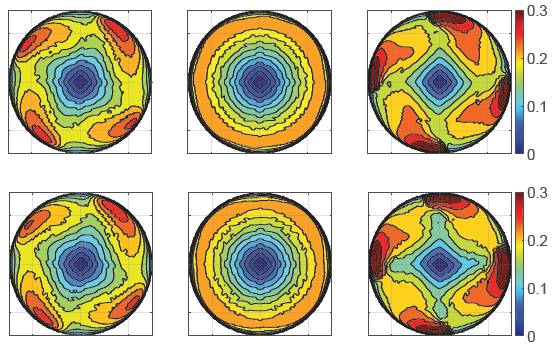}
\put(-190,130){(b)} 
\put(-178,118){RNG~{$k$-$\epsilon$}} 
%\put(-95,118){rea {$k$-$\epsilon$}} 
\put(-125,118){realizable {$k$-$\epsilon$}} 
\put(-55,118){{$k$-$\omega$}~SST}  
\put(-195,75){{\rotatebox{90}{short}}} 
\put(-195,15){{\rotatebox{90}{long}}} 
\vspace{0.02\textwidth}
\includegraphics[height=0.30
\textwidth, trim=0cm 0cm 9cm 0cm, clip=true] {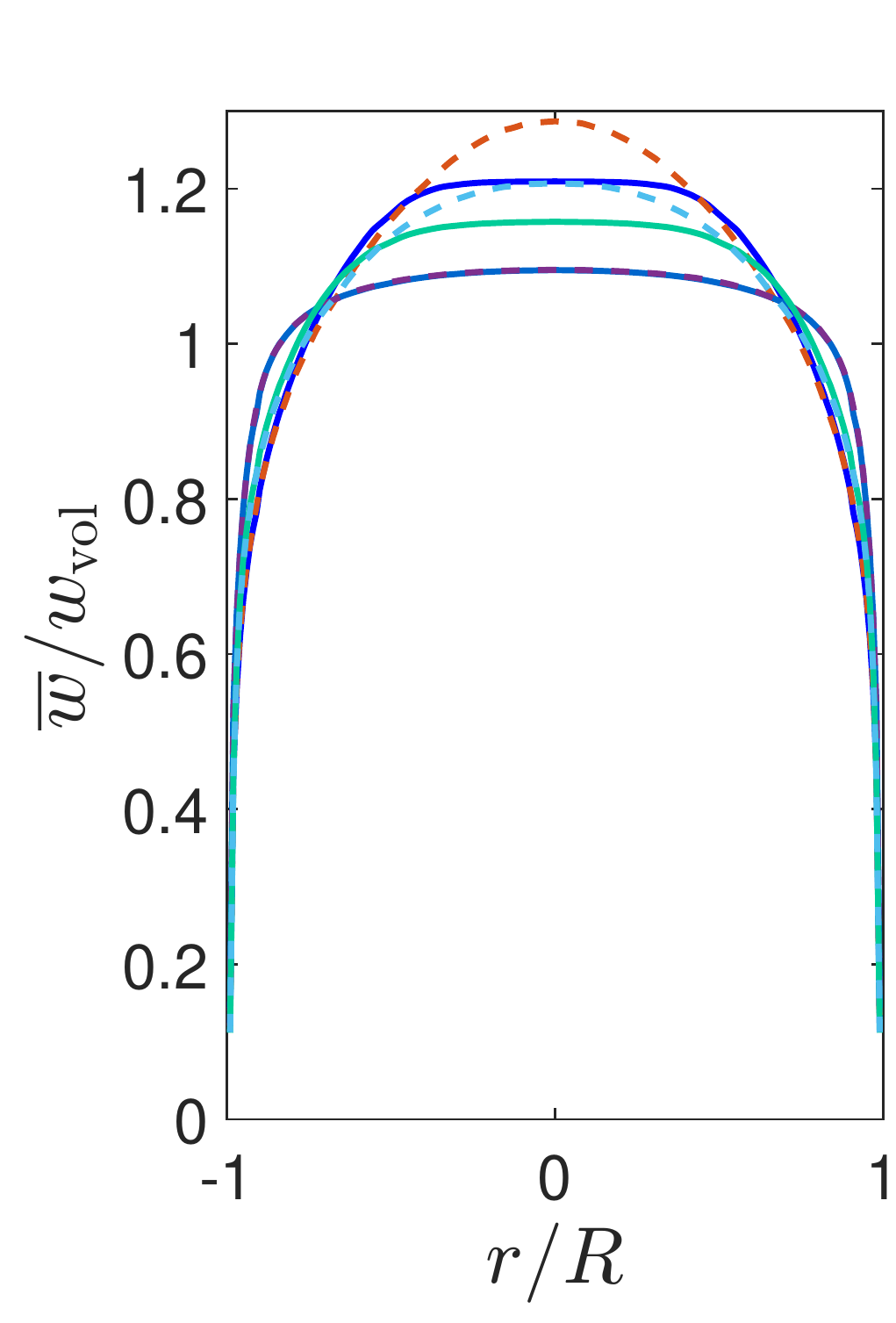}\put(-75,135){(c)} 
\hspace{0.05cm}
\includegraphics[height=0.30
\textwidth, trim=2.4cm 0cm 9cm 0cm, clip=true] {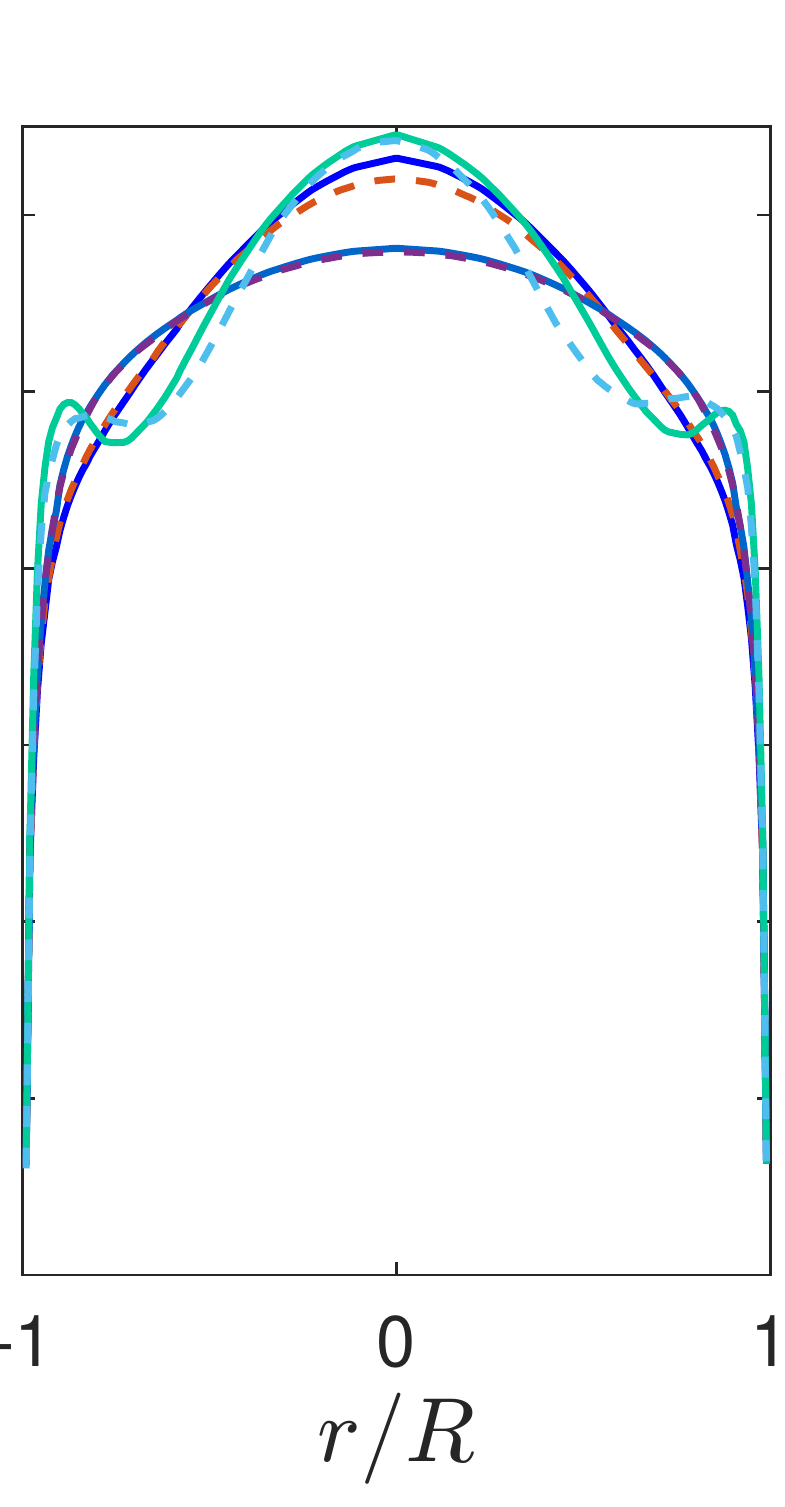}\put(-75,135){(d)} 
\includegraphics[height=0.30
\textwidth, trim=9cm 0cm 0cm 0cm, clip=true] {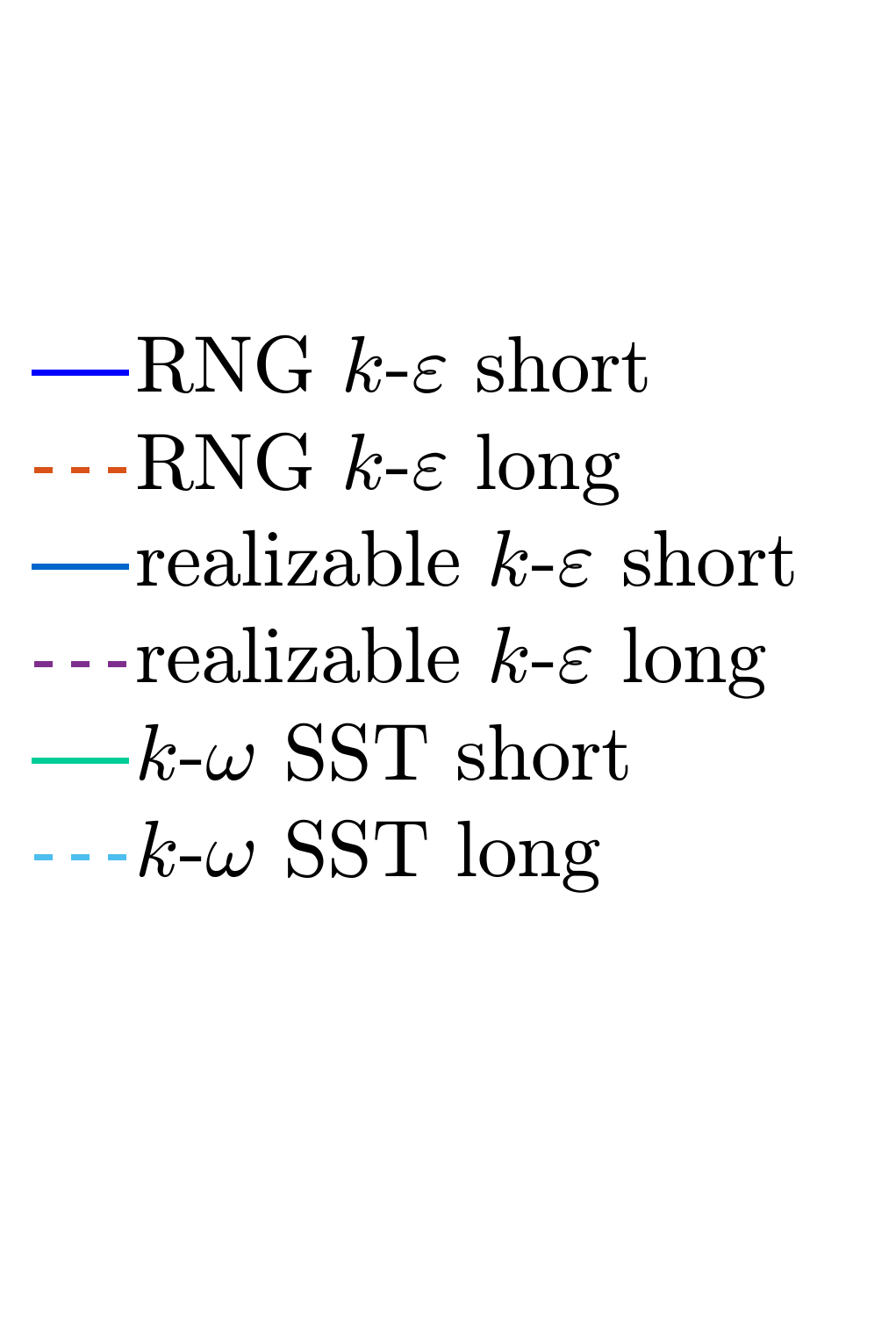}
\hspace{-0.2cm}
\includegraphics[height=0.30
\textwidth, trim=0cm 0cm 9cm 0cm, clip=true] {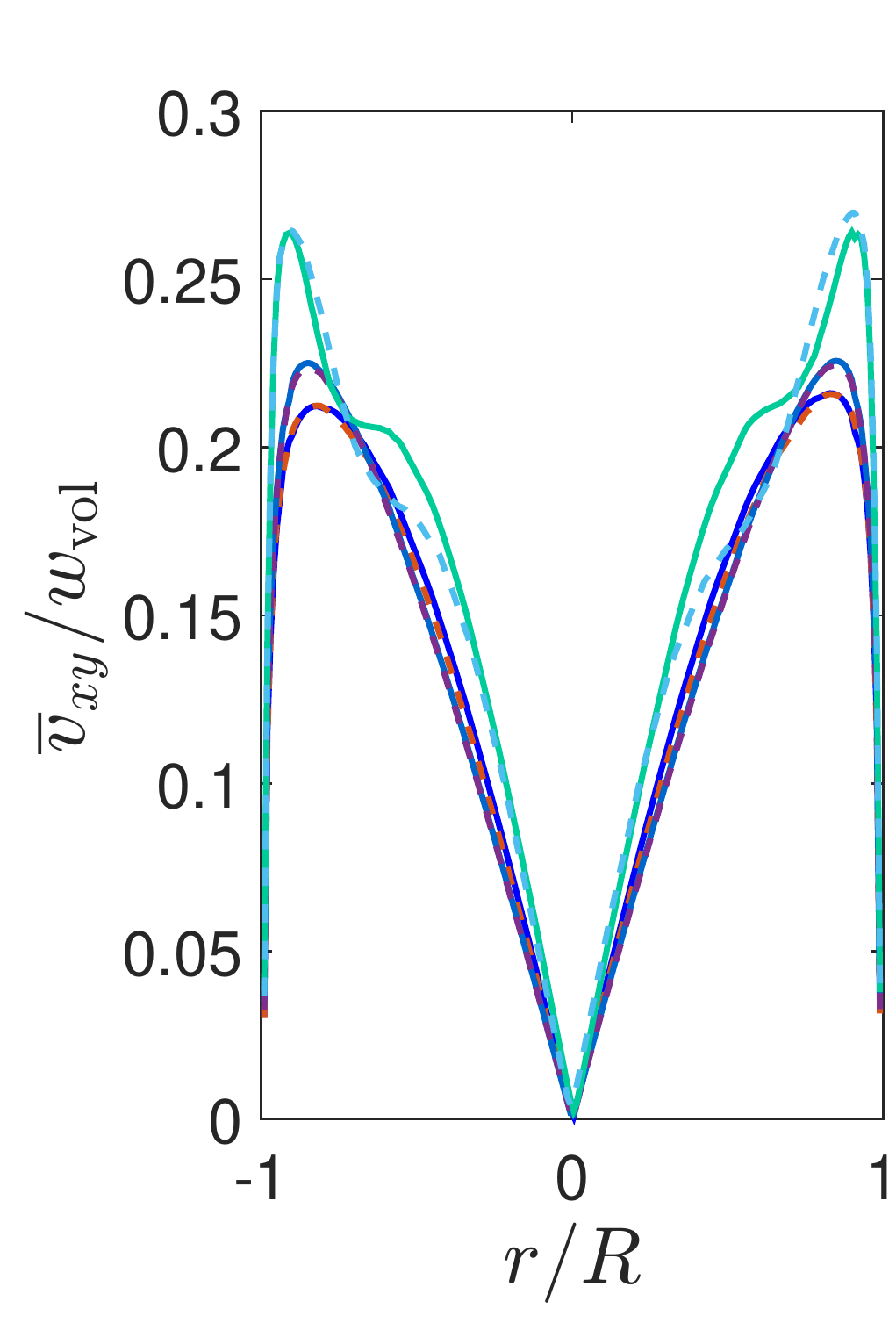}\put(-75,135){(e)} 
\hspace{0.0cm}
\caption{The length of the computational domain has minor influence on the downstream solution: Comparison of $\overline w/w_{\rm vol}$ and $\overline v_{xy}/w_{\rm vol}$ for the long and short computational domains for three different turbulence models. (a) Contour plots of $\overline w/w_{\rm vol}$ at $12.0D$ downstream. (b) Contour plots of $\overline v_{xy}/w_{\rm vol}$ at $12.0D$ downstream.  (c) Averaged $\overline w/w_{\rm vol}$ profile at $2.0D$ upstream. (d) Averaged $\overline w/w_{\rm vol}$  profile at $12.0D$ downstream. (e) Averaged $\overline v_{xy}/w_{\rm vol}$ profile at $12.0D$ downstream. Note that the secondary flow upstream from the swirl generator is zero.}
\label{fig:longComp}
\end{figure*}
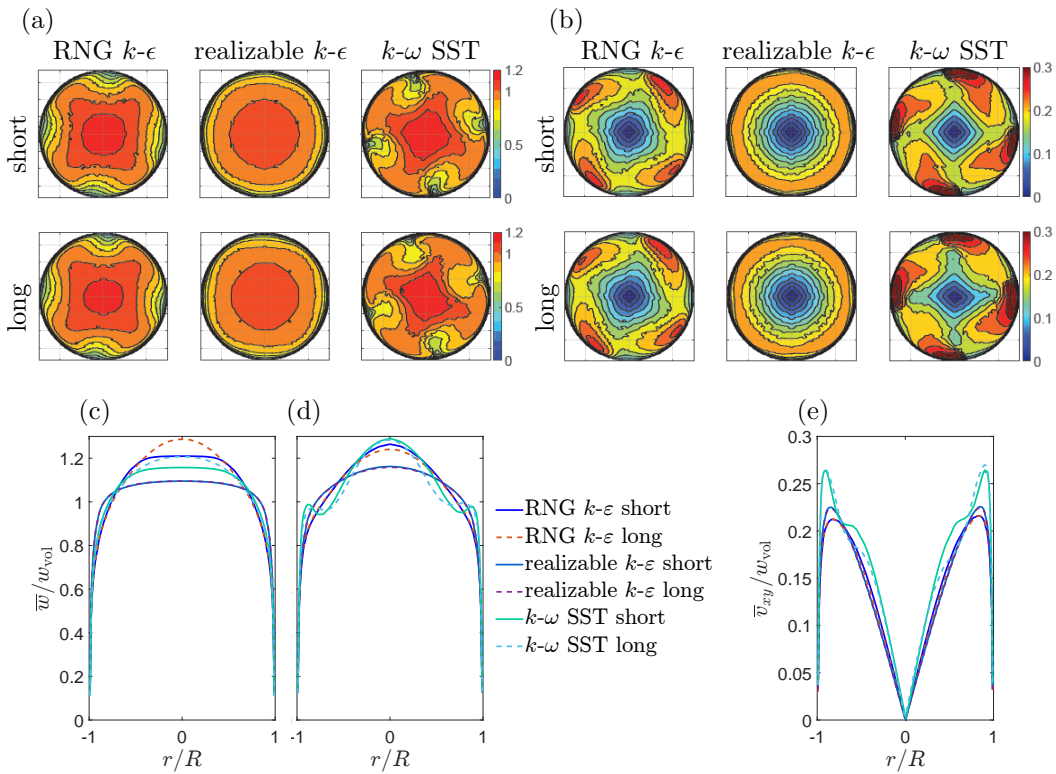
%%%%%%%%%%%%%%%

%%%%%%%%%%%%%%%
\begin{figure*}[t!]
\centering
%\graphicspath{ {./pictures//} }
$\quad$
\includegraphics[width=0.9\textwidth] {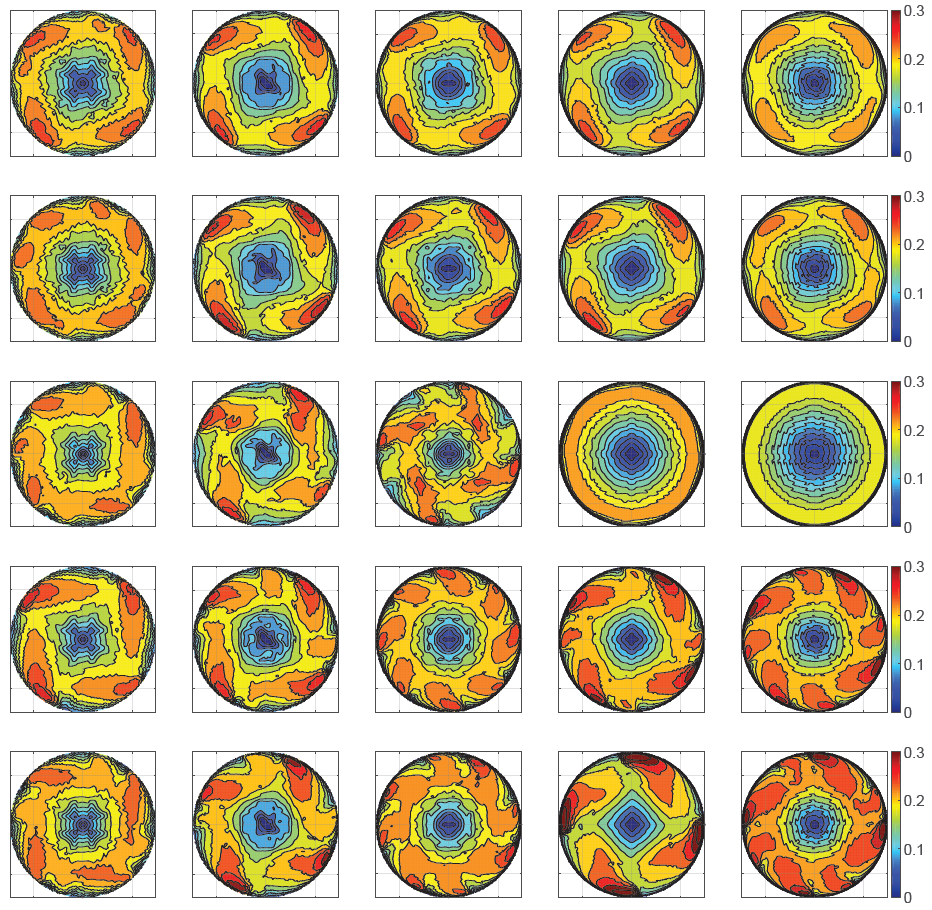}
\put(-395,410){$\sim 10^6$}
\put(-320,410){$\sim 2\cdot 10^6$} 
\put(-235,410){$\sim 4\cdot 10^6$} 
\put(-155,410){$\sim 10\cdot 10^6$} 
\put(-70,410){$\sim 20\cdot 10^6$} 
%\put(-220,330){amount of cells} 
%\put(-430, 410){\rotatebox{90}{std {$k$-$\epsilon$}}}
\put(-430, 350){\rotatebox{90}{standard {$k$-$\epsilon$}}}
\put(-430,270){\rotatebox{90}{RNG~{$k$-$\epsilon$}}} 
%\put(-430,220){\rotatebox{90}{rea {$k$-$\epsilon$}}}
\put(-430,180){\rotatebox{90}{realizable {$k$-$\epsilon$}}} 
\put(-430, 120){\rotatebox{90}{$k$-$\omega$}} 
\put(-430, 20){\rotatebox{90}{{$k$-$\omega$}~SST}} 
\caption{Mesh dependency of different turbulence models: Comparison of $\overline v_{xy}/w_{\rm vol}$ at $12.0D$ predicted by various models and mesh sizes.} %$12.0D$ downstream of the swirl disturbance generator
\label{fig:all12D}
\end{figure*}
\subsection{Impact of the mesh size and domain length}
\label{sec:mesh}
%%%%%%%%%%%%%%%

To identify an optimal trade-off between simulation time and accuracy of results, we assess the impact of the mesh size and the impact of the length of the computational domain on the solution of different turbulence models. To that end, we run all turbulence models with five different mesh sizes. To quantify the effect of the domain length, we first perform simulations with a $20.0D$ inlet and outlet corresponding to a total domain length of $800.0 \, \rm{mm}$. In Figure~\ref{fig:longComp}, we compare results of both inlet lengths for three selected turbulence models. Panel (a) of Figure~\ref{fig:longComp} shows the axial component $\overline{w}$ of the velocity $12.0D$ downstream from the swirl disturbance generator. Further we visualize the magnitude of the secondary flow
\begin{equation}\label{eq:v_xy}
\overline v_{xy}=\sqrt{\overline u^2+\overline v^2} ,
\end{equation}
as shown in panel (b) of Figure~\ref{fig:longComp}. Visual comparison of both results shows that the characteristic flow patterns are maintained across different domain lengths although there are small-scale differences in the velocity field. To assess these differences in more detail, we compare velocity profiles over $r/R$ of each case upstream and downstream from the swirl generator. We compute averages of 10 individual linear profiles with equidistant angular spacing of $18^\circ$ in analogy with the experimental measurement grid shown in Figure~\ref{fig:swirlGrid}~(b). The comparison of the averaged axial profiles upstream from the swirl generator shows that both RNG~{$k$-$\epsilon$} and {$k$-$\omega$}~SST models provide flow profiles closer to the fully developed reference with the longer inlet, while the realizable {$k$-$\epsilon$} model appears to provide similar results for both inlet lengths (Figure~\ref{fig:longComp}~(c)). In the cross-section $12.0D$ downstream from the disturbance generator, the RNG~{$k$-$\epsilon$} model shows a slight attenuation of the axial peak velocity for the long inlet and the {$k$-$\omega$}~SST model exhibits a smoother profile for the long inlet. The corresponding $\overline v_{xy}$ profiles are shown in panel (e) of Figure~\ref{fig:longComp}. Both the realizable {$k$-$\epsilon$} model and the RNG~{$k$-$\epsilon$} model show almost identical $\overline v_{xy}$ for the long and the short computational domains. For different lengths of the computational domain, the {$k$-$\omega$}~SST model preserves main features including peaks and gradients close to the wall and the pipe center, but shows small differences in the $\overline v_{xy}$ profile for intermediate $r/R$. These results suggest that the impact of the domain length in the considered cases is minor. Further, this comparison illustrates, that the profile upstream from the swirl disturbance generator has a minor influence on the flow patterns downstream, suggesting that upstream disturbances do not influence the characteristic patterns of the swirling flow field downstream. Therefore, the study is performed using the shorter computational domain, which allows a reduction of the number of cells and the computational time.

%%%%%%%%%%%%%%%
\begin{figure}[t!]
\centering
%\graphicspath{ {./pictures//} }
$\qquad$
\includegraphics[height=0.25
\textwidth, trim=0cm 0cm 7cm 0cm, clip=true] {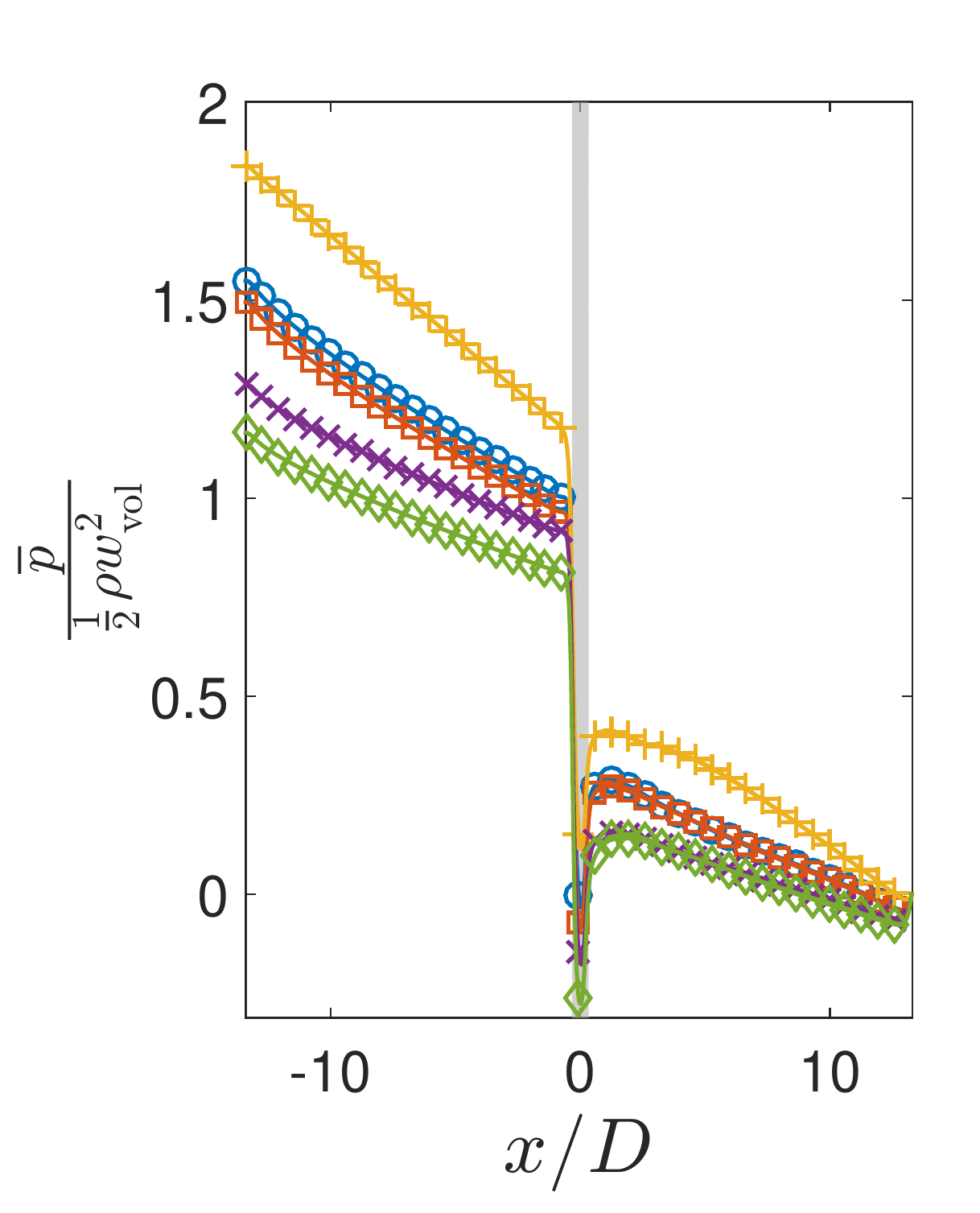}\put(-85,115){(a)} 
\includegraphics[height=0.25
\textwidth, trim=0cm 0cm 0cm 0cm, clip=true] {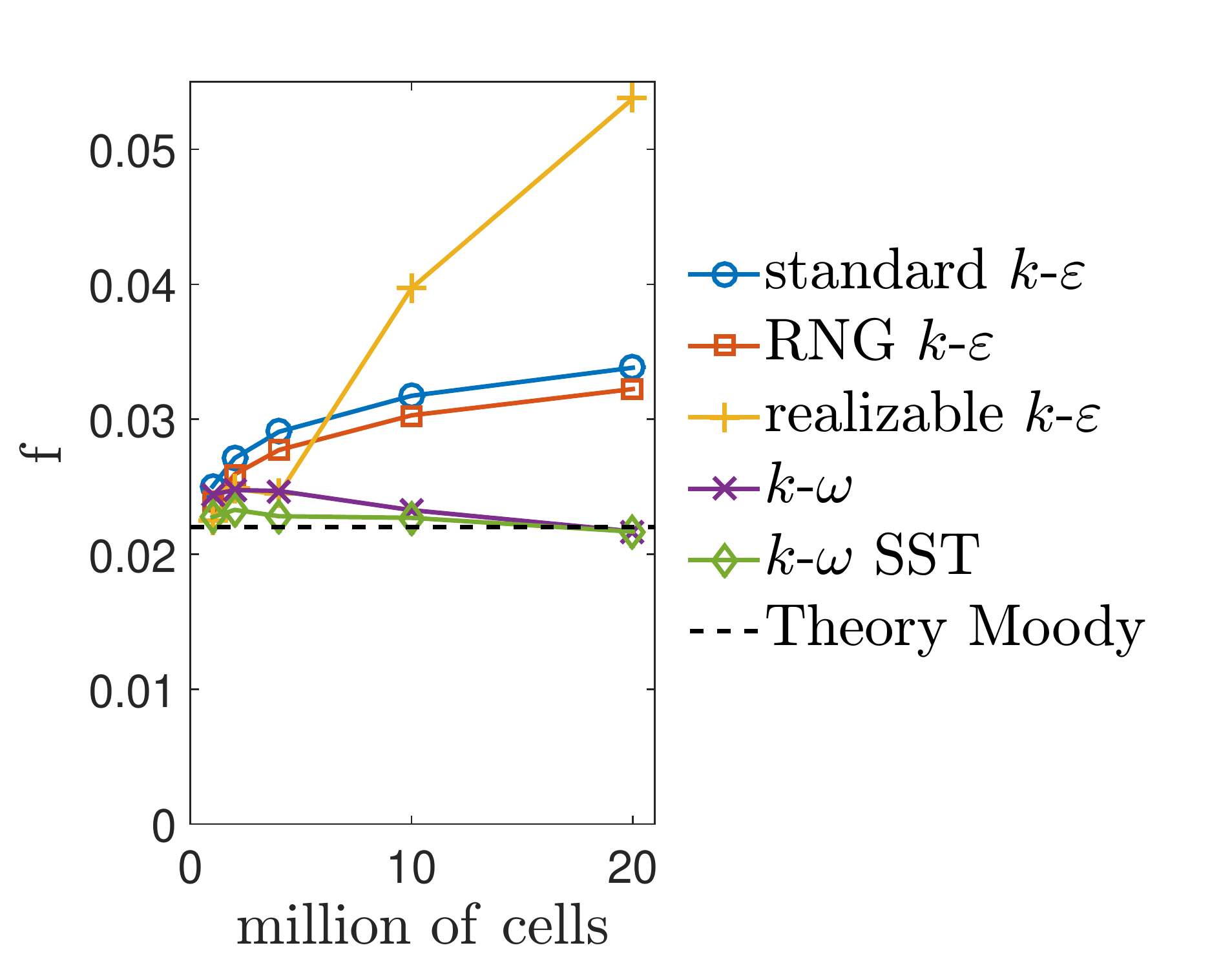}\put(-140,115){(b)} 
\caption{(a) Pressure along the centerline for the different turbulence models. (b) Darcy friction factor for the pipe section between $-3.0D$ and $-1.0D$.  The inlet is located at $x/D\approx-13.0$, the swirl disturbance generator is located at $x/D=0.0$ (shaded area), and the outlet is located at $x/D\approx13.0$.}
\label{fig:pressure}
\end{figure}

Next, we study the influence of the mesh-size on the solution. 
Figure~\ref{fig:all12D} shows the secondary flow predicted by different turbulence models and different meshes. The standard {$k$-$\epsilon$} model and the RNG~{$k$-$\epsilon$} model show little dependency on the mesh size, maintaining similar flow patterns across different mesh sizes. In contrast, the realizable {$k$-$\epsilon$} model exhibits a significant difference between solutions on coarser and finer meshes. While the coarser meshes show small-scale flow patterns, the finer meshes predict a homogeneous swirling flow without characteristic small-scale patterns. In general, the models from the {$k$-$\epsilon$} family tend to predict smaller magnitudes of $\overline v_{xy}$ with increasing mesh sizes, whereas models from the {$k$-$\omega$} family tend to predict higher magnitudes of $\overline v_{xy}$ with increasing mesh size. Under the assumption of monotonic convergence, this suggests that swirling flows computed with models from the {$k$-$\epsilon$} family should be interpreted as a lower estimate for the secondary flow, whereas results computed with models from the {$k$-$\omega$} family should be interpreted as an upper estimate for the secondary flow. 

Further, we study the convergence of the pressure drop with increasing mesh-size. The pressure along the center-line of the pipe is shown in Figure~\ref{fig:pressure} (a) for all considered turbulence models. As expected, there is a significant pressure drop across the swirl disturbance generator, while the pressure loss is approximately linear with respect to the $z$-coordinate for the pipe sections before and after the swirl disturbance generator. To compare the pressure loss in the straight pipe with theoretical references, we compute the Darcy--Weisbach friction factor 
\begin{equation}\label{eq:darcy}
f= \frac{2D}{\rho_w Lw_{\rm vol}^2}\Delta \overline p,
\end{equation}
where $\rho$ is the density of water, $L$ is the associated length of the pipe-section (from $-3.0D$ to $-1.0D$, to consider the section where the flow is most developed), and $\Delta \overline p$ is the associated pressure loss. The friction factor is computed for all turbulence models and compared to the theoretical reference obtained from the Moody diagram~\cite{Moody1944}. Figure~\ref{fig:pressure} (b) shows that the pressure drop over the straight pipe downstream from the swirl disturbance generator is overpredicted by all considered turbulence models when compared to the theoretical reference from the Moody Diagram. The standrad {$k$-$\epsilon$} model and the RNG {$k$-$\epsilon$} model appear to converge to a value that is higher than the theoretical reference as the mesh size increases. Moreover the realizable {$k$-$\epsilon$} model shows a significant overprediction, where values for the finer meshes do not agree with the theory nor with the other models. Conversely, models from the {$k$-$\omega$} family show a very close agreement with the theory and appear to converge towards the theoretical value for increasing mesh size. For the discussion in the remainder of the article, we use the results obtained with a mesh size of
$20\cdot 10^6$ cells.

%%%%%%%%%%%%%%%
\subsection{Velocity fields}
\label{sec:vel}
%%%%%%%%%%%%%%%

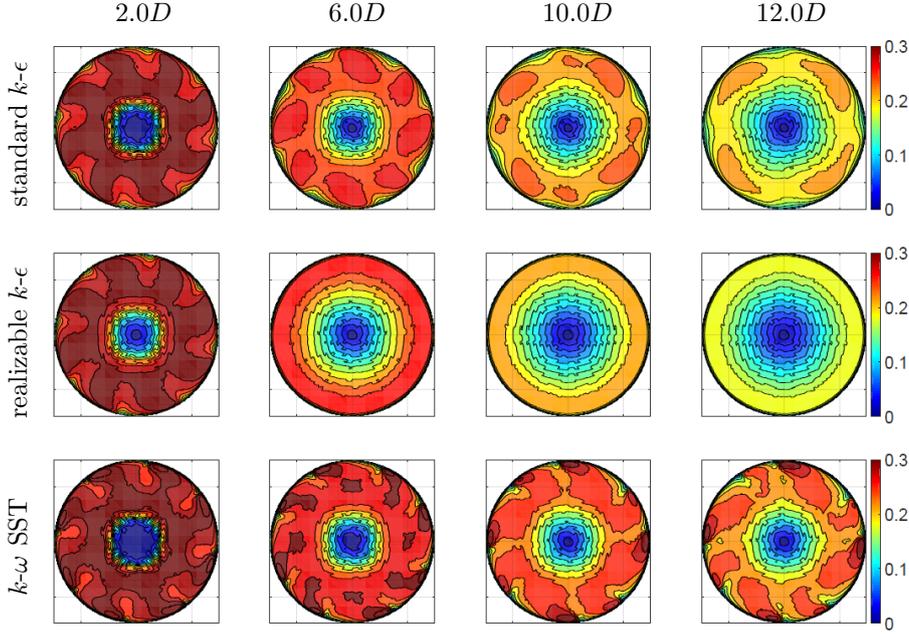
\begin{figure*}[t!]
\centering
%\graphicspath{ {./pictures//} }

\includegraphics[width=0.7\textwidth] {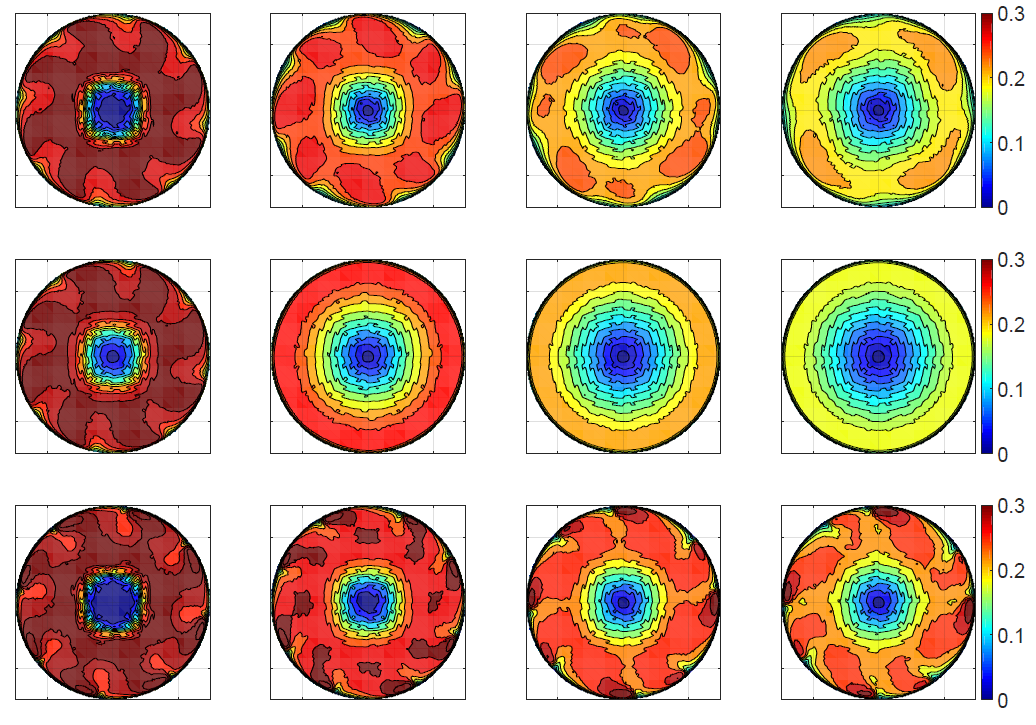}
\put(-300, 235){$2.0D$} 
\put(-220, 235){$6.0D$} 
\put(-140, 235){$10.0D$} 
\put(-60, 235){$12.0D$} 
%\put(-290, 150){\rotatebox{90}{std {$k$-$\epsilon$}}} 
\put(-340, 165){\rotatebox{90}{standard {$k$-$\epsilon$}}} 
%\put(-290,85){\rotatebox{90}{rea {$k$-$\epsilon$}}} 
\put(-340,85){\rotatebox{90}{realizable {$k$-$\epsilon$}}} 
\put(-340, 18){\rotatebox{90}{{$k$-$\omega$}~SST}} 

\caption{Evolution of the secondary flow downstream of the swirl generator: Contour plots of $\overline v_{xy}/w_{\rm vol}$ for various cross-sections downstream.}
\label{fig:downS}
\end{figure*}

\begin{figure}[t!]
\centering
%\graphicspath{ {./pictures//} }

$\qquad\qquad$
\includegraphics[width=0.4\textwidth, trim=0cm 9cm 0cm 10cm, clip=true] {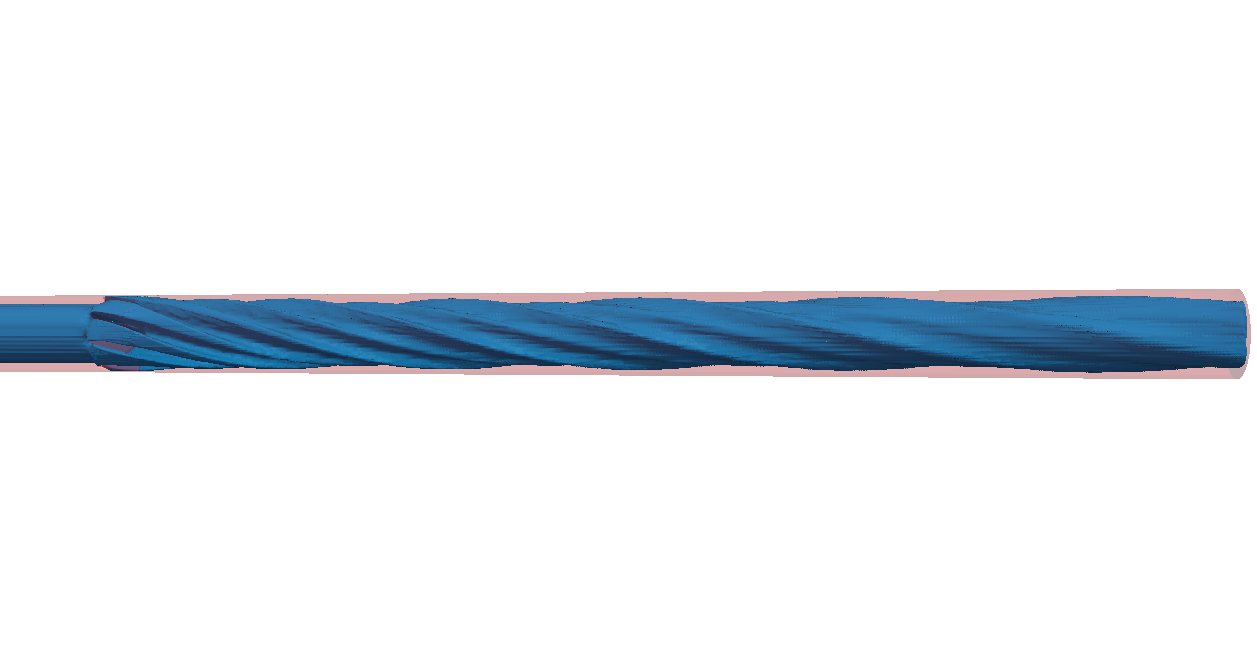}\put(-190, 22){standard {$k$-$\epsilon$}} 
\vspace{0.0cm}
$\qquad\qquad$
\includegraphics[width=0.4\textwidth, trim=0cm 9cm 0cm 10cm, clip=true] {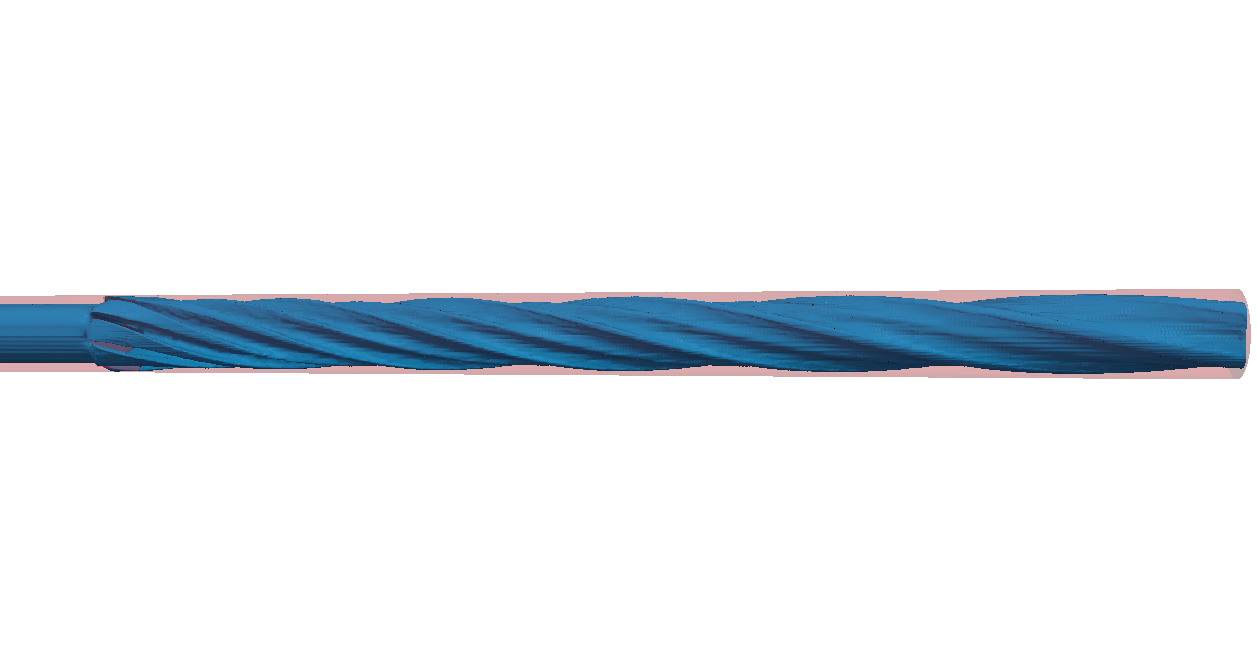}\put(-190, 22){RNG~{$k$-$\epsilon$}} 
\vspace{0.0cm}
$\qquad\qquad$
\includegraphics[width=0.4\textwidth, trim=0cm 9cm 0cm 10cm, clip=true] {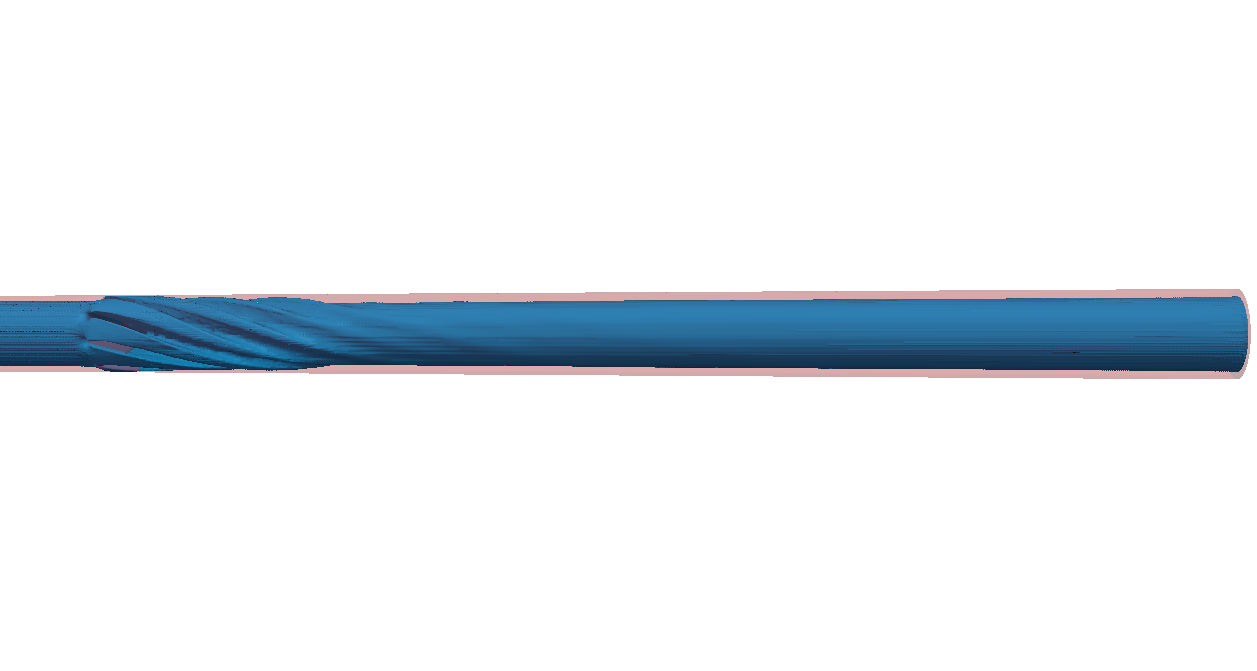}\put(-190, 22){realizable {$k$-$\epsilon$}} 
\vspace{0.0cm}
$\qquad\qquad$
\includegraphics[width=0.4\textwidth, trim=0cm 9cm 0cm 10cm, clip=true] {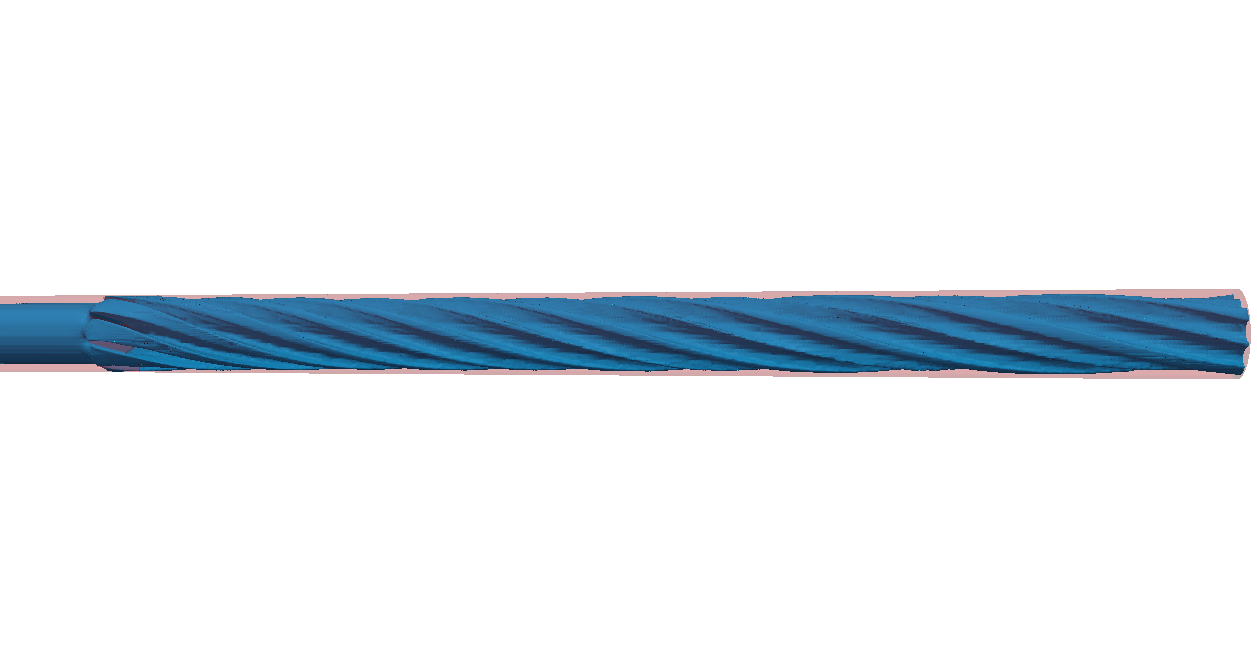}\put(-190, 22){{$k$-$\omega$}} 
\vspace{0.0cm}
$\qquad\qquad$
\includegraphics[width=0.4\textwidth, trim=0cm 9cm 0cm 10cm, clip=true] {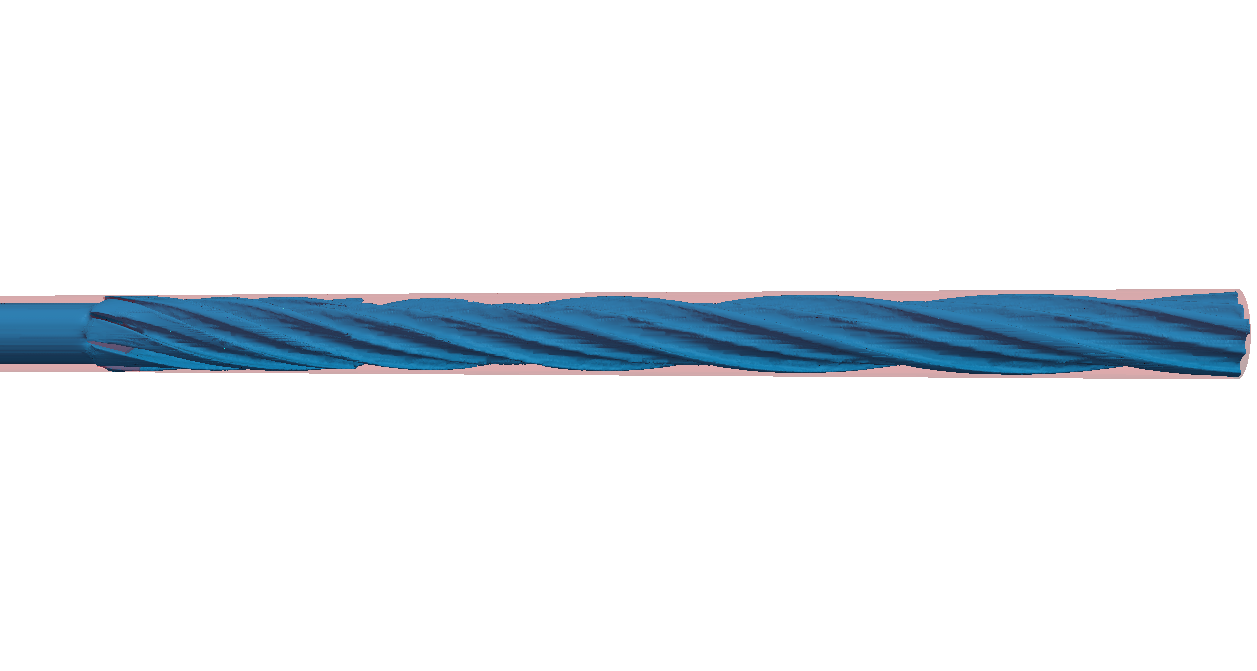}\put(-190, 22){{$k$-$\omega$}~SST} 
\caption{Isosurfaces of $\overline w =w_{vol}$ representing the evolution of the flow downstream of the swirl generator for all the considered turbulence models.}
\label{fig:isoS}
\end{figure}

\begin{figure*}[t!]
\centering
%\graphicspath{ {./pictures//} }
%\includegraphics[width=1\textwidth] {Second_flow}%\put(-220,150){(a)} 
\includegraphics[width=0.6\textwidth] {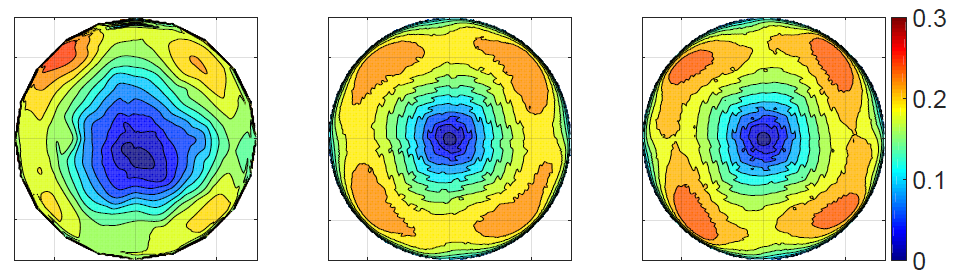}
\put(-280, 80){(a) experimental}
\put(-182, 80){(b) standard {$k$-$\epsilon$}}
\put(-85, 80){(c) RNG~{$k$-$\epsilon$}} 
\vspace{0.2cm}
\includegraphics[width=0.6\textwidth] {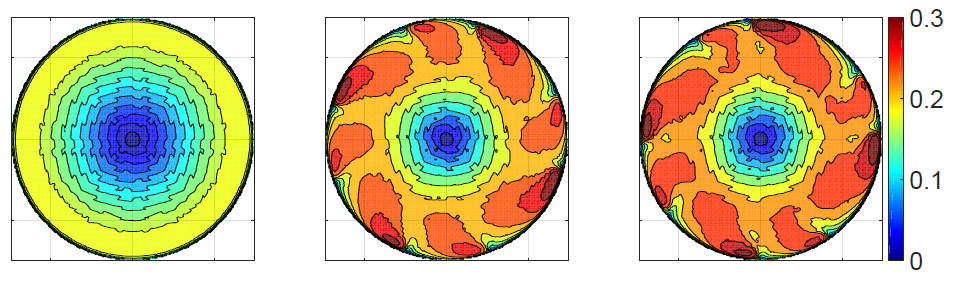}
\put(-280, 80){(d) realizable {$k$-$\epsilon$}}
\put(-165, 80){(e) {$k$-$\omega$}} 
\put(-83, 80){(f) {$k$-$\omega$}~SST}
\caption{Comparison of simulations and experiments $12.0D$ downstream from the disturbance swirl generator: $\overline v_{xy}/w_{\rm vol}$ from LDV experiments (a) and predicted by the different turbulence models (b)--(f).}
\label{fig:simVSexp}
\end{figure*}
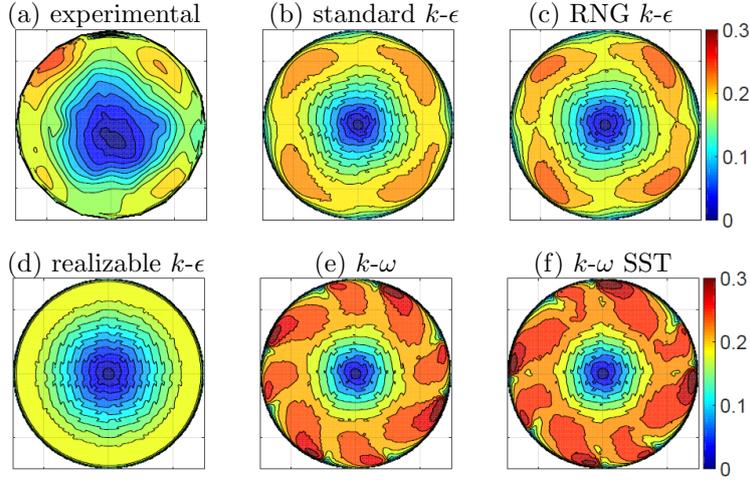

For the comparison of flow patterns at different cross-sections downstream of the swirl disturbance generator, we focus on three of the models since the structures predicted by the RNG~{$k$-$\epsilon$}, the standard {$k$-$\epsilon$} models, and both {$k$-$\omega$} models are found to be qualitatively similar (see, for example, Figure~\ref{fig:simVSexp}). The flow patterns of the secondary flow predicted at four locations are shown in Figure~\ref{fig:downS} and isosurfaces corresponding to $\overline w=w_{vol}$ of the domain downstream of the swirl disturbance generator are shown in Figure~\ref{fig:isoS}.

All considered models predict eight characteristic flow patterns at the cross-section $2.0D$ downstream, which are associated with the eight blades of the swirl disturbance generator. At $6.0D$, the realizable {$k$-$\epsilon$} model does not show any small-scale flow patterns in $\overline v_{xy}$, but a radially homogeneous swirling flow. In contrast, the {$k$-$\omega$}~SST model and the {$k$-$\epsilon$} model still predict eight patterns with elevated $\overline v_{xy}$. The homogeneous swirling flow of the realizable {$k$-$\epsilon$} model is maintained further downstream with a continuous reduction of $\overline v_{xy}$. For the {$k$-$\epsilon$} model, the initial structures start dissipating and merging at around $10.0D$, resulting in only four visible flow structures at the $12.0D$ measurement cross-section. The {$k$-$\omega$}~SST model maintains the eight characteristic flow patterns throughout the entire simulation domain, although at the $12.0D$ measurement section the dissipation of four of them starts to become visible.

Next, we compare the simulations with the experimental results at the section $12.0D$ downstream of the swirl disturbance generator. Figure~\ref{fig:simVSexp} shows the secondary flow obtained from simulations along with the corresponding experimental results. Visual inspection of the contour plots suggests that the standard {$k$-$\epsilon$} model and the RNG~{$k$-$\epsilon$} model provide the best approximations of the experimentally observed four characteristic patterns in the outer region of the pipe. Conversely, the realizable {$k$-$\epsilon$} model provides a more realistic approximation of the velocity magnitude close to the center of the pipe.

%%%%%%%%%%%%%%%
%%%%%%%%%%%%%%%

%%%%%%%%%%%%%%%
%%%%%%%%%%%%%%%

%%%%%%%%%%%%%%%%
%\begin{figure*}[h!]
%\centering
%\graphicspath{ {./pictures//} }
%%\includegraphics[width=1\textwidth] {Second_flow}%\put(-220,150){(a)} 
%\includegraphics[height=0.25\textwidth, trim=1cm 0cm 4cm 0cm, clip=true] {vel_xymagnitude_2DpaperFormat}\put(-80, 115){(a) experimental}
%\hspace{0.2cm}
%\includegraphics[height=0.25\textwidth, trim=1cm 0cm 4cm 0cm, clip=true] {00_kEpsilon20Mcells_secondFlow_plane_12Dpdf} \put(-80, 115){(b) standard {$k$-$\epsilon$}}
%\hspace{0.2cm}
%\includegraphics[height=0.25\textwidth, trim=1cm 0cm 0cm 0cm, clip=true] {00_RNG20Mcells_secondFlow_plane_12Dpdf}\put(-100,115){(c) RNG~{$k$-$\epsilon$}} 
%
%\includegraphics[height=0.25\textwidth, trim=1cm 0cm 4cm 0cm, clip=true] {00_realisableKEp20Mcells_secondFlow_plane_12Dpdf}\put(-85,115){(d) realizable {$k$-$\epsilon$}}
%\hspace{0.2cm}
%\includegraphics[height=0.25\textwidth, trim=1cm 0cm 4cm 0cm, clip=true] {00_komega20Mcells_secondFlow_plane_12Dpdf}\put(-65, 115){(e) {$k$-$\omega$}}  
%\hspace{0.2cm}
%\includegraphics[height=0.25\textwidth, trim=1cm 0cm 0cm 0cm, clip=true] {kOmegaSST_20Mcells_secondFlow_plane_12Dpdf}\put(-100, 115){(f) {$k$-$\omega$}~SST}
%
%\caption{Comparison of simulations and experiments $12.0D$ after the disturbance swirl generator: in-plane flow components measured with LDV experiments (a) and predicted by the different turbulence models (b)--(f).}
%\label{fig:simVSexp}
%\end{figure*}
%%%%%%%%%%%%%%%%

\begin{figure*}[t!]
\centering
%\graphicspath{ {./pictures//} }
\includegraphics[height=0.25
\textwidth, trim=0cm 0cm 6cm 1cm, clip=true] {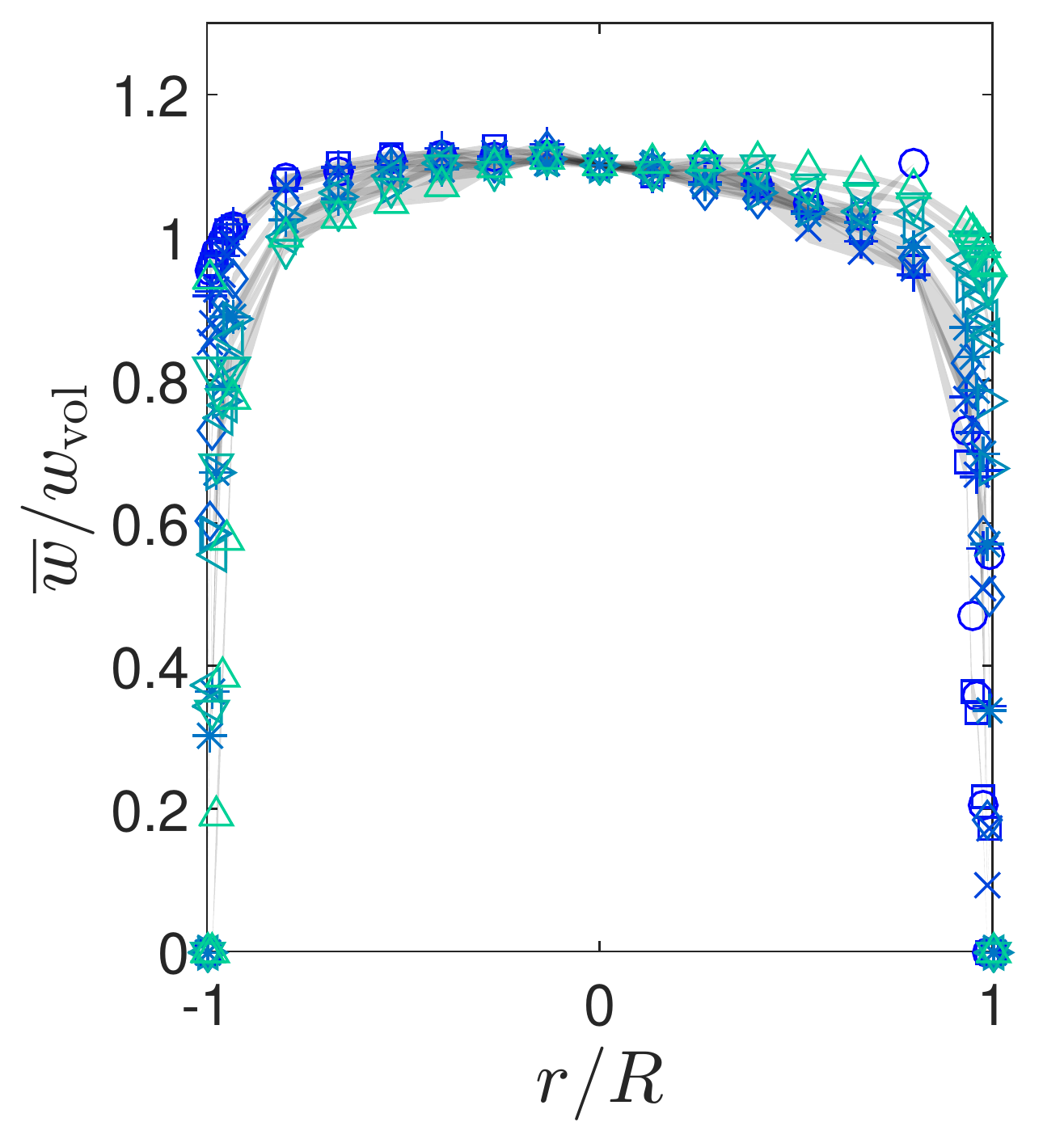}\put(-90,120){(a)} 
\hspace{0.05cm}
\includegraphics[height=0.25
\textwidth, trim=0cm 0cm 6cm 1cm, clip=true] {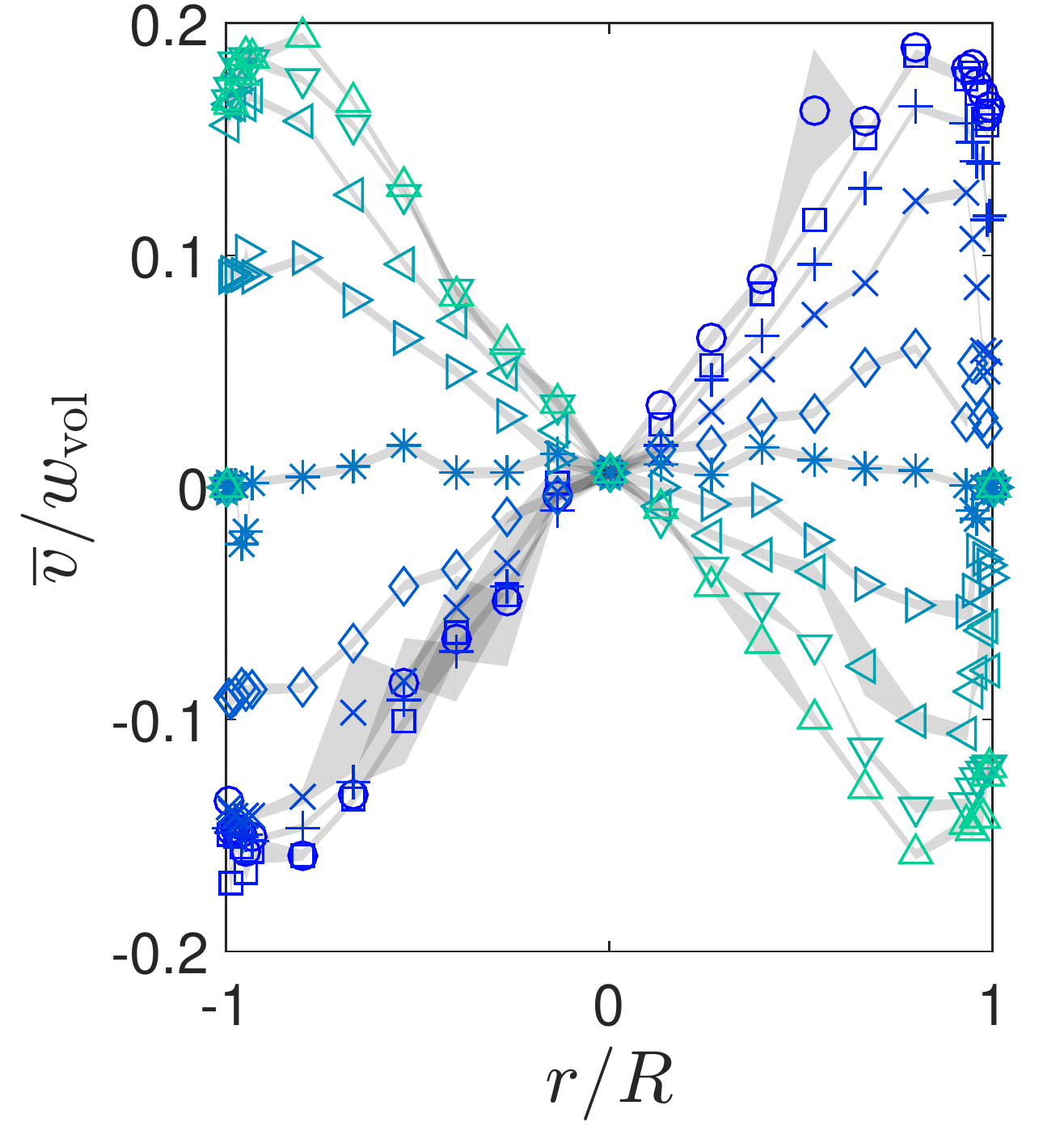}\put(-90,120){(b)} 
\hspace{0.05cm}
\includegraphics[height=0.25
\textwidth, trim=0cm 0cm 0cm 1cm, clip=true] {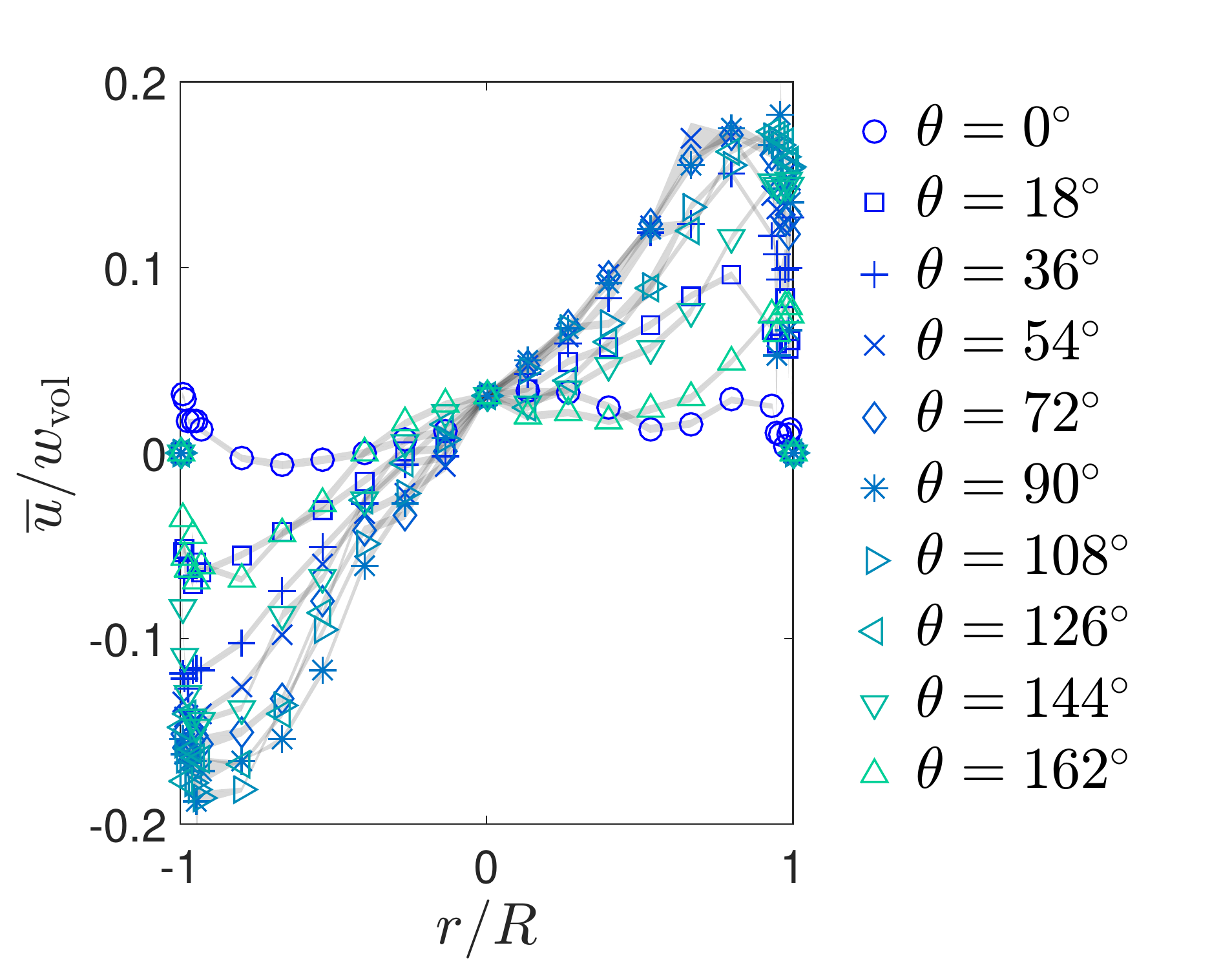}\put(-138,120){(c)}
\caption{Experimental velocity profiles at $12.0D$ downstream for the three Cartesian components with standard errors (shaded area): (a) $z$-component $\overline w/w_{\rm vol}$, (b) $y$-component $\overline v/w_{\rm vol}$, and (c) $x$-component $\overline u/w_{\rm vol}$.}
\label{fig:expProfiles}
\end{figure*}

\begin{figure*}[t!]
\centering
%\graphicspath{ {./pictures//} }

\includegraphics[height=0.25
\textwidth, trim=0cm 0cm 6.2cm 1cm, clip=true] {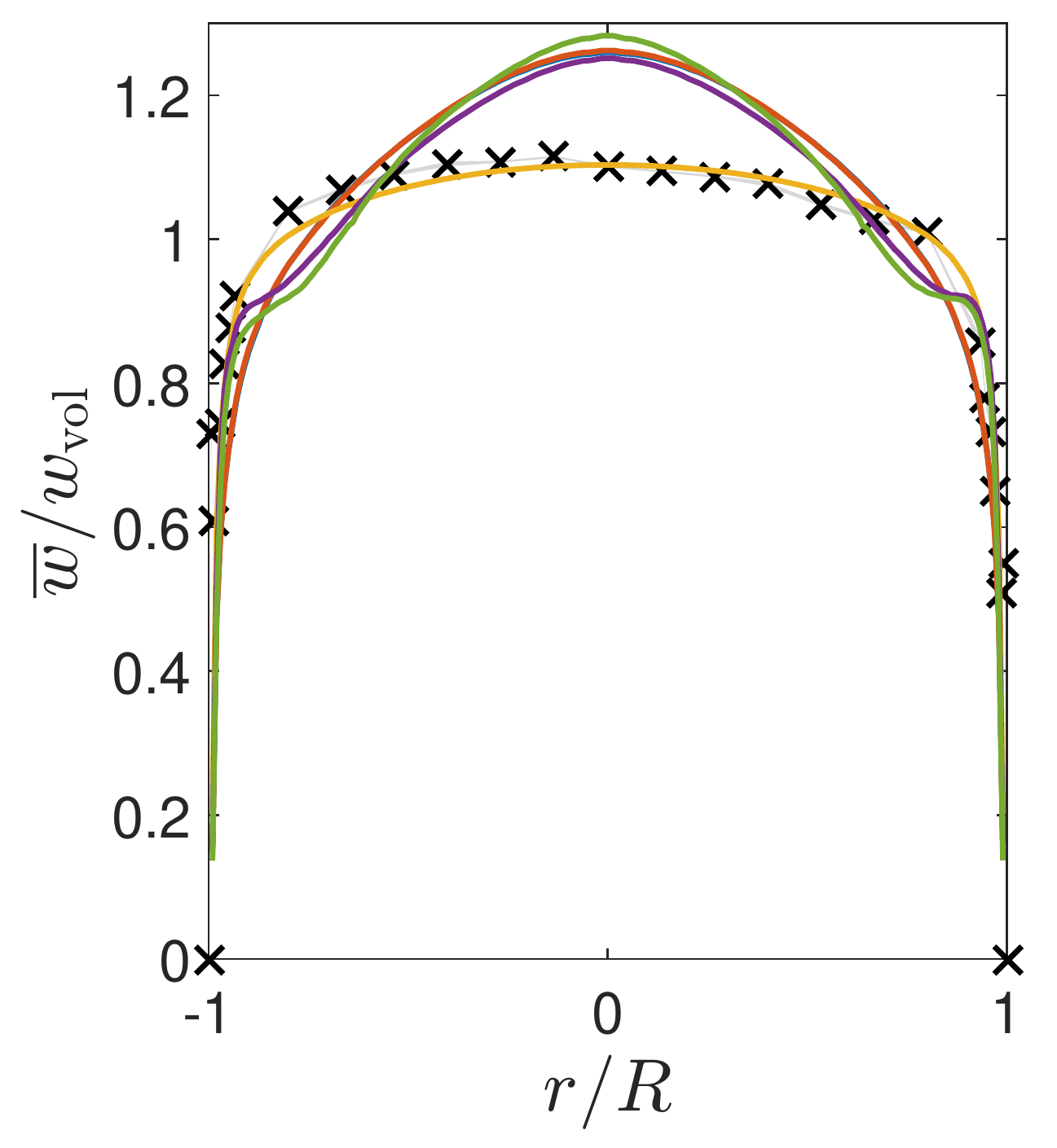}\put(-90,120){(a)}
\hspace{0.1cm}
\includegraphics[height=0.25
\textwidth, trim=0cm 0cm 6.2cm 1cm, clip=true] {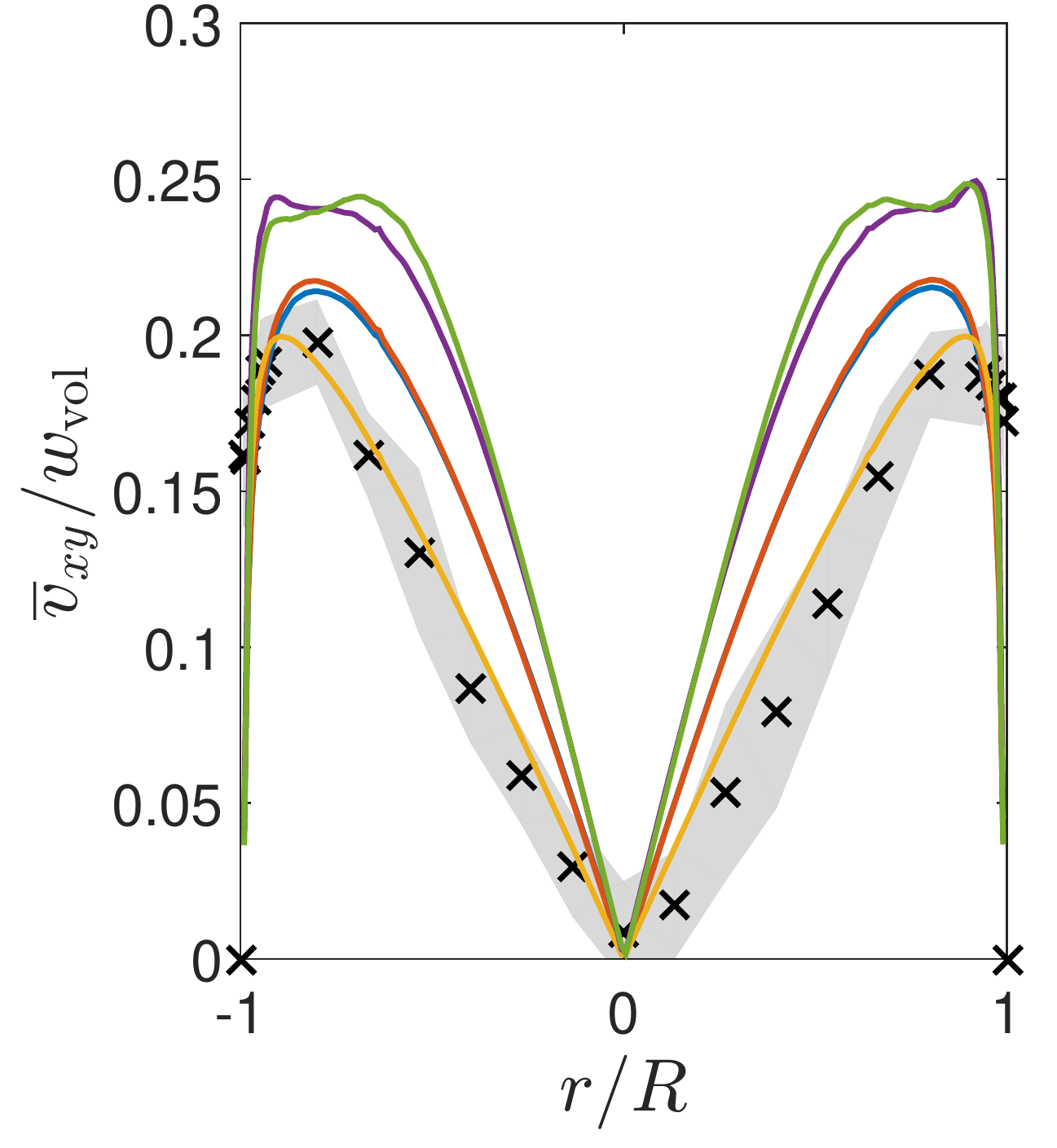}\put(-90,120){(b)}
\includegraphics[height=0.25
\textwidth, trim=11cm 0cm 0cm 1cm, clip=true] {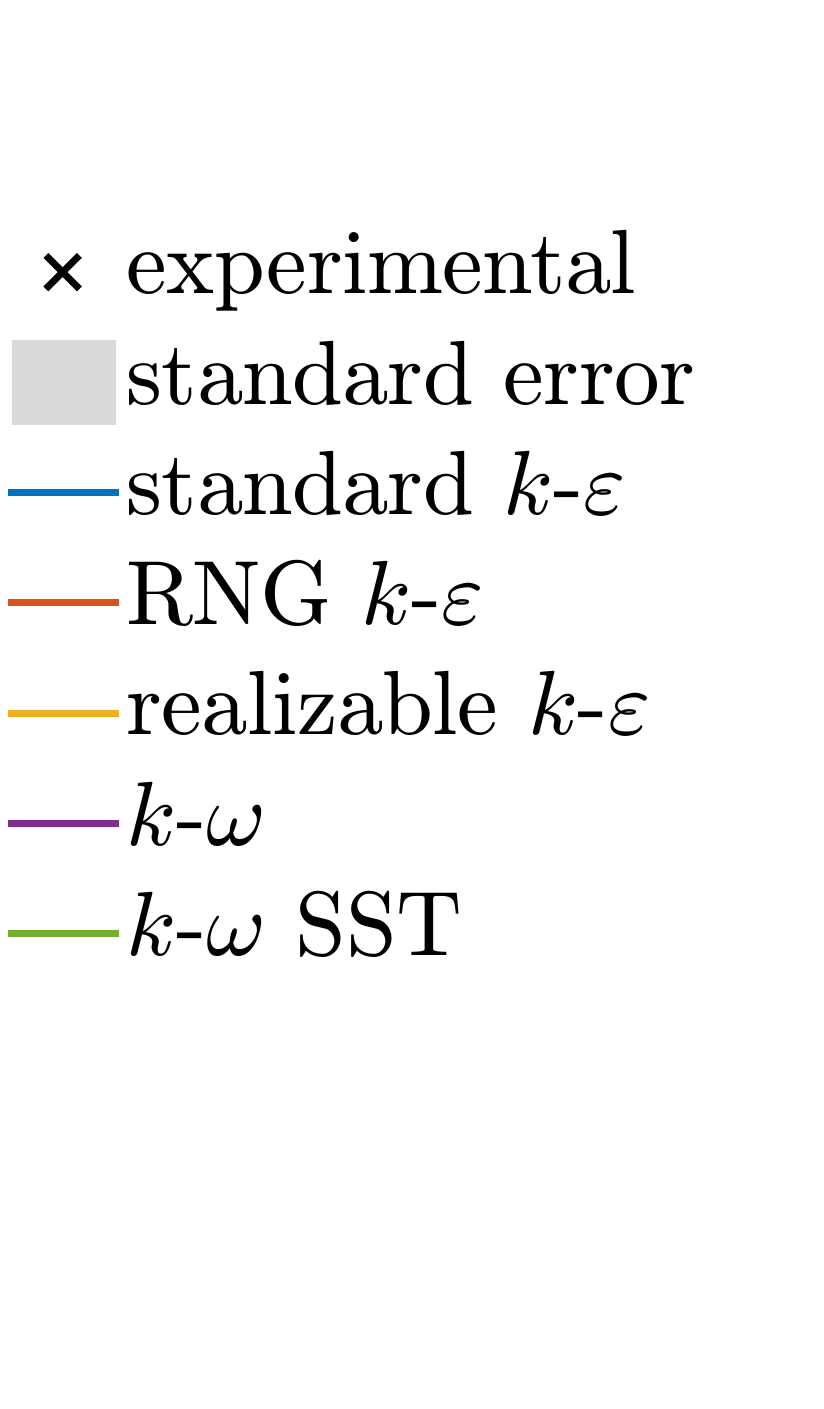} % This is the legend
\caption{Comparison of simulations and experiments $12.0D$ downstream from the swirl disturbance generator: (a) Averaged profiles of $\overline w/w_{\rm vol}$ profile and associated standard error~\ref{eq:AVG_SE1}. (b) Averaged $\overline v_{xy}/w_{\rm vol}$ profile and associated standard error~\ref{eq:AVG_SE2}.}
\label{fig:shadedComp}
\end{figure*}

For the comparison of individual profile paths, we use 10 averaged profiles as described in Section~\ref{sec:mesh}. The individual measured velocity profiles are shown in Figure \ref{fig:expProfiles} along with the corresponding standard error. The standard errors associated with the averaged profiles are computed through error propagation calculus using the standard error~\ref{eq:sdErr} of individual profiles. For the averaged axial velocity profile, we compute the associated standard error 
\begin{equation} \label{eq:AVG_SE1}
\overline \sigma_{\overline w} = \frac{1}{m}\left(\sum\limits_{i=1}^{m}\sigma_{\overline w,i}^2 \right)^{1/2},
\end{equation}
where $m=10$ is the number of individual profiles.
Similarly, for the secondary flow, assuming uncorrelated $\sigma_{\overline u}$ and $\sigma_{\overline v}$, we find
\begin{equation} \label{eq:AVG_SE2}
\overline \sigma_{xy} = \frac{1}{m}\left(
						\sum\limits_{i=1}^{m} \frac
												{\overline u_i^2 \sigma_{\overline u,i}^2+\overline v_i^2 \sigma_{\overline v,i}^2}
												{\overline u_i^2+\overline v_i^2}
										\right)^{1/2}
					.
\end{equation}
The averaged profiles are shown in Figure~\ref{fig:shadedComp}. 
The shaded area indicates the associated standard error of the experimental results with a coverage factor $\kappa=2.0$. The profiles of $\overline w$ indicate that the realizable {$k$-$\epsilon$} model is the only model that does not predict a spurious pronounced velocity peak at the pipe center, but an almost flat plateau-like profile that closely approximates the experimental data (Figure~\ref{fig:shadedComp}~(a)).

Similarly, the profiles of $\overline v_{xy}$ show that the near-field flow produced by the standardized swirl disturbance generator is similar to solid-body rotation (Figure~\ref{fig:shadedComp}~(b)). Importantly, the flow field generated by the standardized swirl disturbance generator according to {EN~ISO~4064-2:2014}~\cite{ENSO4064} and {OIML~R~49-2:2013}~\cite{OIML2013} appears to lack a transition of the flow from solid body rotation to Rankine-vortex type flow as reported by Vaidya et al.~\cite{Vaidya2011} for a swirling flow generated by a rotating honeycomb. All turbulence models appear to overpredict $\overline v_{xy}$ and, as previously seen in Figure~\ref{fig:simVSexp}, the highest values are given by the models of the {$k$-$\omega$} family. Overall, the best prediction of the experimental data is provided by the realizable {$k$-$\epsilon$} model which appears to capture the mean flow velocity more accurately than the other considered models.

%%%%%%%%%%%%%%%
%%%%%%%%%%%%%%%

%%%%%%%%%%%%%%%
%%%%%%%%%%%%%%%

%%%%%%%%%%%%%%%
\subsection{Performance indicators}
\label{sec:Pfact}
%%%%%%%%%%%%%%%

Performance indicators are useful integral metrics to quantify flow conditions. Following Yeh and Mattingly~\cite{Yeh1994} and M\"{u}ller and Dues~\cite{Muller2007}, we compute the swirl angle $\phi$, the profile factor $K_{\rm p}$, the asymmetry factor $K_{\rm a}$, and the turbulence factor $K_{\rm Tu}$. A detailed definition of the performance indicators is provided in~\ref{sec:defPerfParam}.

In Table~\ref{t:perFact}, we compare performance indicators from computational and experimental results as well as available experimental results of Eichler and Lederer~\cite{Eichler2015} with a DN80 pipe diameter and a flow rate of $96.221\, \rm{m^3/h}$ (${\rm Re}\approx 4.0\cdot10^5$). The values of performance indicators are reported as mean values of 10 profiles shown in Figure~\ref{fig:swirlGrid} (b) along with the associated standard deviation. According to the definition~\ref{eq:K_a01}, the asymmetry factor $K_{\rm a}$ is signed. However, to report a meaningful non-zero mean, we average the magnitude of $K_{\rm a}$. For the simulations, the turbulence factor $K_{\rm Tu}$ is computed using the turbulent kinetic energy $k$ that is part of the solution of the corresponding turbulence model, such that
%
%%%%%%%%%%%%%%%
\begin{equation}\label{eq:TuI}
{\rm Tu}=\frac{\sqrt{2k/3}}{|\overline \bfu|^2}.
\end{equation}
%%%%%%%%%%%%%%%
%
Notice that~\ref{eq:TuI} corresponds to a three-dimensional turbulence factor, while the $K_{\rm Tu}$ value of the experiments corresponds to individual axial profiles.

The values in Table~\ref{t:perFact} indicate that the realizable {$k$-$\epsilon$} model provides low standard deviations in performance indicators. This is consistent with the circular symmetry of the predicted flow fields, which results in almost identical individual profiles, as discussed in Section~\ref{sec:vel}. The profile factor $K_{\rm p}$ of the realizable {$k$-$\epsilon$} model shows good agreement with the LDV results, while all other models predict a higher value for this parameter. This substantiates the results discussed in Section~\ref{sec:vel}, where the realizable {$k$-$\epsilon$} model is found to provide the best prediction of the experimental axial profile.

The asymmetry factor $K_{\rm a}$ obtained from the simulations is significantly lower than the experimental value. Consequently, none of the models shows potential to provide an accurate estimation of this parameter. However, the experimental setup is always subject to small imperfections, which results in non-symmetric flow profiles, whereas the computational model is fully symmetric by definition. Further, previous investigations (see, for example, Eichler and Lederer~\cite{Eichler2015}) reported that the swirling flow field may develop instabilities resulting in symmetry breaking at a certain distance downstream. The present computational model does not capture the effect of symmetry breaking. The stability of the swirling flow could be assessed computationally through small perturbations, for example in the boundary conditions. However, such an investigation goes beyond the scope of this article and should be investigated in future studies.

The turbulence factor $K_{\rm Tu}$ is approximated reasonably well by the considered turbulence models. The best prediction is provided by the RNG~{$k$-$\epsilon$} model and the standard {$k$-$\epsilon$} model. The values obatined from the {$k$-$\omega$} model and the {$k$-$\omega$}  SST model are lower than the experimental values. The realizable {$k$-$\epsilon$} model appears to fail in capturing the behaviour of $K_{\rm Tu}$, giving a considerable overprediction.

The swirl angle is well predicted by the three {$k$-$\epsilon$} models, whereas the {$k$-$\omega$} models predict values that exceed those obtained from the LDV experiments. In the present scenario, the realizable {$k$-$\epsilon$} model gives the closest value to the experiments, which supports the accuracy of this model in predicting the experimental radial profiles as discussed in Section~\ref{sec:vel}. In general, simulations with turbulence models from the {$k$-$\epsilon$} family appear to provide the most reasonable predictions of the performance indicators. Yet, none of the considered models gives an accurate estimation of all four parameters.

Comparing our experimental and computational results with the experimental results of Eichler and Lederer~\cite{Eichler2015}, we see a reasonably close agreement for $K_{\rm p}$, $K_{\rm Tu}$, and $\phi$. The observed discrepancy in $K_{\rm a}$ can be influenced by the experimental setup of the LDV system where the pipe center needs to be determined by the operator. This procedure is not very robust and can result in small displacements from the actual center causing asymmetric flow profiles. Further, the conditions for the swirling flow field to become unstable remain elusive, as discussed by Tawackolian~\cite{Tawackolian2013a} and Büker~\cite{Buker2010}. Without additional information, the distinction between actual asymmetries in the flow field and asymmetries due to the uncertainties associated with the determination of the pipe center is inaccessible.

%%%%%%%%%%%%%%%
\begin{table*}[t!]%[H] add [H] placement to break table across pages
\centering
\footnotesize
%\small
\caption{Performance indicators at the cross-section $12.0D$ downstream from the swirl disturbance generator.}
\label{t:perFact}
\begin{tabular}{p{3.5cm}lllp{2cm}p{2cm}lp{2cm}}
\toprule
    				&$K_{\rm p}[-]$ 		&$K_{\rm a}[\%]$		&$K_{\rm Tu}[-]$ 		&$\phi[\rm{deg}]$\\ 
\hline
standard {$k$-$\epsilon$}	&$1.173\pm0.131$ 	& $0.062\pm0.035$ 	& $1.792\pm0.004$ 	& $12.269\pm0.550$\\
RNG~{$k$-$\epsilon$}  	&$1.186\pm0.174$ 	& $0.051\pm0.045$ 	& $1.910\pm0.007$ 	& $12.514\pm0.856$\\
realizable {$k$-$\epsilon$}  	&$0.438\pm0.005$ 	& $0.002\pm0.001$ 	& $3.867\pm0.000$ 	& $11.292\pm0.017$\\
{$k$-$\omega$} 	 	&$1.163\pm0.152$ 	& $0.241\pm0.325$ 	& $1.176\pm0.004$ 	& $15.068\pm1.062$\\
{$k$-$\omega$}~SST  	&$1.360\pm0.207$ 	& $0.309\pm0.356$ 	& $1.340\pm0.006$ 	& $15.293\pm1.521$\\ 
\hline
Present experiments 	&$0.470\pm0.196$ 	& $0.975\pm1.120$ 	& $2.071\pm0.054$ 	& $11.426\pm0.561$\\ 
\hline
Experiments $13.0D$ downstream with DN80 and $Re\approx 4\cdot10^5$~\cite{Eichler2015}  	& 0.61 			& 0.26 			& 2.14 			&13.0\\ 

\bottomrule

\end{tabular}
\end{table*}
%%%%%%%%%%%%%%%

%%%%%%%%%%%%%%%
\section{Discussion of model performance and modeling assumptions}
\label{sec:Pfact}
%%%%%%%%%%%%%%%
%
In conclusion, eddy-viscosity models only provide limited accuracy for reliable predictions of swirling flow. However, these models are still dominant in industrial applications since other methods such as large eddy simulations (LES) still have a significant computational cost, which is presently not achievable in the industrial time-scale.

To discuss how different modeling assumptions of the considered eddy-viscosity models influence the predictions of swirling pipe-flow, we take the standard {$k$-$\epsilon$} model as a baseline reference.

The RNG~{$k$-$\epsilon$} model includes an extra term in the $\epsilon$ equation that is expected to enhance its performance in flows with high strain rate and streamline curvature (see, for example, Escue and Cui~\cite{Escue2010}). Although there is a considerable streamline curvature in the present scenario, we do not detect major improvements with respect to the standard {$k$-$\epsilon$} model. However, a potential improvement of the RNG~{$k$-$\epsilon$} model by using an empirical modification of the turbulent viscosity is part of various commercial codes and was applied by Escue and Cui~\cite{Escue2010} and Saqr et al.~\cite{Saqr2012}. The implementation of such a modification in OpenFOAM might well yield further improvements, but goes beyond the scope of this article.

Similarly, the realizable {$k$-$\epsilon$} model aims to provide improvements with respect to the standard {$k$-$\epsilon$} model in swirling flows through taking into consideration the effects of mean rotation on computing the turbulent viscosity. In view of this property, the ability of the realizable {$k$-$\epsilon$} model to capture the velocity magnitudes more accurately than the standard {$k$-$\epsilon$} model appears to be consistent with the modeling assumptions.

The {$k$-$\omega$} model by Wilcox~\cite{Wilcox1988} is designed to improve the performance of the standard {$k$-$\epsilon$} model in the viscous layer of the near-wall regions and in adverse pressure gradients, promising more accurate results for free shear flows and separated flows. Additionally, the {$k$-$\omega$}~SST model combines the {$k$-$\omega$} model and the standard {$k$-$\epsilon$} model through a blending function that prompts the use of the {$k$-$\omega$} model in the near wall regions and the {$k$-$\epsilon$} model in regions far from the wall. The {$k$-$\omega$}~SST model was originally formulated by Menter~\cite{Menter1994} with the goal of modeling aerodynamics flows. Menter and Esch~\cite{Menter2001} presented a modification of the {$k$-$\omega$}~SST model which includes a new definition of the eddy viscosity using the strain rate instead of the vorticity. This modification is aimed at extending the applicability of the model to a wider range of flows. Yet, in the present scenario, we see no improvements on using the {$k$-$\omega$} model or the {$k$-$\omega$}~SST model rather than the standard {$k$-$\epsilon$} model. With respect to the experimental results, both models give the most unsatisfactory results for predicting realistic velocity profiles in the $12.0D$ measurement plane. Conversely, we find that the {$k$-$\omega$} model and the {$k$-$\omega$}~SST model provide realistic values of the pressure loss for sufficiently fine meshes. Further, the modeling assumptions suggest that the {$k$-$\omega$} models are likely to provide a realistic prediction of the flow in the pipe section close the swirl disturbance generator and inside the swirl disturbance generator, were the influence of the wall region becomes more relevant. Yet, these modeling assumptions render the {$k$-$\omega$} models deficient for predicting the evolution of the flow further downstream, where the influence of the wall is less dominant.

%%%%%%%%%%%%%%%
\section{Conclusions}
\label{sec:Conclusions}
%%%%%%%%%%%%%%%

We assessed the potential of five different turbulence models to predict flow patterns downstream of a standardized swirl disturbance generator with different meshes and mesh topologies. The numerical results were validated through a systematic comparison with LDV experiments. For sufficiently fine meshes, the realizable {$k$-$\epsilon$} model gives the most reasonable prediction of both axial and in-plane velocity profiles, when compared to experimental results. However, the realizable {$k$-$\epsilon$} model fails to predict the characteristic small-scale flow patterns determined experimentally, which are well-captured by the standard {$k$-$\epsilon$} model and the RNG~{$k$-$\epsilon$} model. The pressure drop over the straight pipe-section before the swirl generator is overpredicted by models from the {$k$-$\epsilon$} family when compared to a theoretical reference determined from the Moody diagram. In particular the realizable {$k$-$\epsilon$} model appears to provide elevated values for the friction factors. On the other hand, the {$k$-$\omega$} models predict values that are closest to the theoretical reference. We conclude that, in general, {$k$-$\epsilon$} models appear to provide more accurate predictions of the velocity fields but fail to provide realistic predictions of the pressure drop. Conversely, {$k$-$\omega$} models give a better prediction for the pressure drop. Overall, the performance of the investigated eddy-viscosity models appears to be limited, which is in agreement with expectations (see, for example, Jakirli\'c et al.~\cite{Jakirlic2002}). However, the usage of these models still reflects current industry practice and this detailed study including experimental validation facilitates a more targeted model selection and guidance for the interpretation of results.
Additionally, a comparison of the present results with other references with different pipe diameters and flow rates shows that the performance indicators exhibit similar values. This suggests that, despite small differences in the swirl disturber geometries, the flow fields generated at different pipe diameters and flow rates are approximately self-similar. Consequently, small differences in the geometry of the swirl disturbance generators have only secondary effects on the flow field.

%%%%%%%%%%%%%%%%
\appendix
%%%%%%%%%%%%%%%%%%%%%%%%%%%%%%%%%%%%%%%

\section{Definition of performance indicators}
\label{sec:defPerfParam}
%%%%%%%%%%%%%%%%%%%%%%%%%%%%%%%%%%%%%%
\subsection{Profile factor}
Following Yeh and Mattingly~\cite{Yeh1994}, the dimensionless profile factor $K_p$ is defined as
\begin{equation}\label{eq:K_p01}
K_p = \frac{K_{p, m}}{K_{p, s}},
\end{equation}
with
\begin{equation}
K_{p, m}=\frac{1}{w_{\rm vol}D} \int\limits_{-R}^{R} (w_m - \overline w) \, {\rm{d}}r
\end{equation}
and
\begin{equation}
K_{p, s}=\frac{1}{w_{\rm vol}D} \int\limits_{-R}^{R} (w_{m,s} - w_s) \, {\rm{d}}r,
\end{equation}
where $w_m=\overline w(r=0)$ is the velocity at the pipe center, $w_{m,s}$ is the velocity of the norm profile at the pipe center, and $w_s$ is the velocity of the norm profile. The profile factor is a measure for peakness ($K_p>1$) or flatness ($K_p<1$) of measured velocity profiles with respect to standard profiles such as Hagen--Pouiseuille for laminar flow or Gersten and Herwig~\cite{Gersten1992,Gersten2005} for turbulent flow. 

%%%%%%%%%%%%%%%%%%%%%%%%%%%%%%%%%%%%%%
\subsection{Asymmetry factor}
Following Yeh and Mattingly~\cite{Yeh1994}, the asymmetry factor 
\begin{equation}\label{eq:K_a01}
K_a= \frac{1}{D} \frac{ \int\limits_{-R}^{R} r \overline w \, {\rm{d}}r}{\int\limits_{-R}^{R} \overline w \, {\rm{d}}r }
\end{equation}
quantifies the relative radial displacement of the center of gravity of the area under the flow profile with respect to the pipe center. 

%%%%%%%%%%%%%%%%%%%%%%%%%%%%%%%%%%%%%%
\subsection{Turbulence factor}

Each LDV point measurement is a collection of a large number of bursts resulting in a histogram (or probability density function) for the axial velocity component. The level of dispersion (i.e. the standard deviation) of this histogram quantifies the turbulence intensity~\ref{eq:Tu01}. As discussed by Durst et al.~\cite{Durst1998} and generalized by Pashtrapanska~\cite{Pashtrapanska2004}, the turbulence intensity~\ref{eq:Tu01} in the core region $-0.2\leq r/R \leq 0.2$ can be estimated as 
\begin{equation}\label{eq:TuEst01}
{\rm{Tu}}_s = 0.13 \left({\rm{Re}} \frac{w_{m,s}}{w_{\rm vol}} \right)^{-1/8} 
\end{equation}
for  ${\rm{Re}} \frac{w_{m,s}}{w_{\rm vol}} \geq 4500$. The turbulence factor $K_{\rm{Tu}}$ is defined as
\begin{equation}\label{eq:K_Tu01}
K_{\rm{Tu}}= \frac{{\rm{Tu}}_{\rm max} }{{\rm{Tu}}_s},
\end{equation}
where ${\rm{Tu}}_{\rm max}$ is the maximum of~\ref{eq:Tu01} in the core region $-0.2\leq r/R \leq 0.2$.

%%%%%%%%%%%%%%%%%%%%%%%%%%%%%%%%%%%%%%
\subsection{Swirl angle}

Following Yeh and Mattingly~\cite{Yeh1995}, the level of swirl can be measured quantitatively through the maximal swirl angle
\begin{equation}\label{eq:swirlAngle}
\phi = \arctan \left ( \frac{|\overline v_{xy}|_{\rm max}}{w_{\rm vol}} \right ).
\end{equation}
However, the precise defition of the swirl angle may vary slightly depending on the author. Geometrically, the swirl angle~\ref{eq:swirlAngle} is the angle between the ideal velocity vector and the actual velocity vector with swirl.

%%%%%%%%%%%%%%%

\section{Mesh quality parameters}
\label{sec:meshQ}
%%%%%%%%%%%%%%%
%
%%%%%%%%%%%%%%%
\begin{table}[h]%[H] add [H] placement to break table across pages
\centering
\footnotesize
%\small
\caption{Mesh quality parameters used in the mesh generation with \emph{snappyHexMesh}.}
\label{t:meshQuality}
\begin{tabular}{llllp{2cm}p{2cm}lp{2cm}}
\toprule
OpenFoam parameter    	&Value \\ 
\hline
maxNonOrtho		&$65.0$ \\
maxBoundarySkewness 	&$20.0$ \\
maxInternalSkewness  	&$4.0$ \\
maxConcave  		&$80.0$ \\
minFlatness			&$0.5$ \\
minVol			  	&$1.0\cdot10^{-20}$ \\ 
minTetQuality		&$1.0\cdot10^{-30}$ \\
minArea 			&$-1$ \\
minTwist  			&$0.05$ \\
minDeterminant		&$0.001$ \\
minFaceWeight	  	&$0.05$ \\ 
minVolRatio			&$0.01$ \\
minTriangleTwist 		&$-1$ \\
nSmoothScale  		&$4$ \\
errorReduction		&$0.75$ \\
\bottomrule

\end{tabular}
\end{table}
%%%%%%%%%%%%%%%

%%%%%%%%%%%%%%%%%%%%%%%%%%%%%%%%%%%%

%%%%%%%%%%%%%%%%%%%%%%%%%%%%%%%
%\bibliography{/Users/dod/Documents/literature/library.bib}
%\bibliographystyle{}
%%%%%%%%%%%%%%%%%%%%%%%%%%%%%%%

\label{lastpage}

\end{document}